# Some Mathematical Issues Pertaining to Translational Gauge Theories

D.H. Delphenich[†]
Physics Department
University of Wisconsin – River Falls
River Falls, WI 54022

*Abstract. Some mathematical aspects of using the translation group as an internal symmetry group in a gauge field theory are presented and discussed. The traditional manner in which gravitation can be accounted for by the introduction of a global frame field on a parallelizable spacetime is reviewed. It is then discussed in the more general context of a global frame field on the bundle of linear frames. In the process, the elements of variational field theory for physical fields defined on G-structures are set down. It is suggested that it is probably more proper to attribute gravitation to a reduction of the bundle of linear frames to {e} – at least over a generic submanifold of spacetime – than to a reduction to the Lorentz group since the Lorentz group is more intrinsic to electromagnetism and gravitation has the character of a "residual" symmetry of spacetime at the astrophysical level.*

**0. Introduction.** One of the most intriguing developments that followed Einstein's theory of gravitation was the observation [**1,2**] that his theory was equivalent to a "translational gauge theory" in which a global frame field and a torsion tensor field that was associated with it played the fundamental role, not a metric tensor field and its Riemannian curvature.

However, in the historical context of these discoveries, numerous details were overlooked simply due to the fact that for the better part of the Twentieth Century all discussion of manifolds and geometry amongst relativity theorists were carried out in a local formulation that was opaque to the global geometrical and topological issues that are now much more apparent to pure mathematicians in the same fields.

In Einstein's theory of gravitation the presence of gravity in spacetime was attributed to the curvature that followed as a consequence of the introduction of a Lorentzian metric $g$ on spacetime. Although the theory is, of course, quite widely accepted by now, nevertheless there are certain aspects of the theory that seem somewhat questionable, or at least puzzling, when viewed in the light of modern field theory, and in particular, gauge field theories. Some examples are:

*a*) There is a crucial difference between defining a local coordinate system $(U, x^\mu)$ about a point of a manifold and defining a local frame field $(U, \mathbf{e}_\mu)$. Although every local coordinate system defines a natural local frame field by way of $\mathbf{e}_\mu = \partial/\partial x^\mu$, the converse is not true; the key issue is one of integrability. As a consequence, one easily confuses demanding the invariance of the field theory under arbitrary coordinate changes – i.e., general covariance – with the invariance of the field theory under arbitrary frame changes.

*b*) The extension of a local frame field to a global one is generally obstructed by the topology of $M$. If such a global frame field exists, i.e., if $M$ is *parallelizable*, then the geometry of $M$ is very closely related to the geometry of Lie groups. Indeed,

---

parallelizable manifolds can be thought of as essentially "almost Lie groups," since Lie groups are examples of parallelizable manifolds and it takes a high degree of symmetry for manifold to be parallelizable. For instance, many homogeneous spaces, such as the two-sphere, are not parallelizable.

*c*) Many of the most essential tools of physical field theories depend upon other local constructions that often *cannot* be globalized except on manifolds that are affine spaces to begin with. In particular, if the action of the translation group $R^n$ on the values of the coordinate functions – viz., $(\varepsilon^\mu, x^\mu) \infty \varepsilon^\mu + x^\mu$ – on a manifold $M$ can be globalized into a transitive effective action on $M$ then $M$ is, by definition, an affine space. If the action is transitive, but not effective, such as the action of R on $S^1$, then $M$ is a more general homogeneous space. This fact is particularly troublesome when trying to introduce the methods of the Fourier transform that are so pervasive in quantum theory.

*d*) When one wishes to translate *frames* on $M$ instead of *points* of $M$, one must introduce a choice of connection on the bundle of linear frames $GL(M)$ and speak of the *parallel* translation of these frames relative to the choice of connection. The Lie algebra of infinitesimal parallel translations on $GL(M)$ – i.e., the Lie algebra of horizontal vector fields under the chosen connection – in not generally Abelian (or even finite-dimensional), unlike $R^n$. Consequently, the action of $R^n$ on $GL(M)$ by parallel translations is not a true action, since the map of $R^n$ into the Lie algebra of $GL(M)$ is not a homomorphism.

However, the reason that one still keeps the translation group in mind in any physical field theory is that in the context of variational field theory translational invariance is closely associated with the conservation of energy-momentum, which is one of the more sacred of first principles in physics.

In this regard, there appears to be a mismatch of infinitesimal symmetries between the geometric left-hand side of Einstein's equation:

$$G_{\mu\nu} = 8\pi\kappa\, T_{\mu\nu}, \tag{0.1}$$

and the dynamical right-hand side. On the left-hand side, one finds the divergenceless Einstein tensor $G_{\mu\nu} = R_{\mu\nu} - 1/2R\, g_{\mu\nu}$ that comes from the Ricci curvature tensor $R_{\mu\nu}$ and the scalar curvature $R$, which are, in turn, derived from the Riemannian curvature tensor $R^\mu_{\nu\rho\sigma}$. Now, at the level of holonomy Riemannian curvature is related to the infinitesimal Lorentz transformations that orthonormal frames undergo during parallel translation. However, the right-hand side of Einstein's equation involves the stress-energy momentum tensor for the matter distribution that serves as the source of the gravitational field. As observed above, according to Nöther's theorem in variational field theory, this tensor appears as a consequence of the invariance of the field-source action under infinitesimal *translations*. This mismatch of infinitesimal symmetries is further compounded when one extends Einstein's theory to the Einstein-Cartan-Sciama-Kibble (ECSK) theory, in which a second field equation couples the torsion tensor of a metric connection to the intrinsic angular momentum tensor; i.e., infinitesimal translations get coupled to infinitesimal Lorentz transformations.

Consequently, one wonders if it is possible to cast the theory of gravitation in a form that makes the role of infinitesimal transformations consistent on both sides. One expects that the appropriate tensor field to be used in the left-hand side would have to be a divergenceless tensor field that one obtains from the torsion tensor field of some connection, which obviously would not be the Levi-Civita connection. In fact, this is exactly what the translational gauge theory of gravitation, which is also referred as *teleparallelism*, promises.

In what follows, we shall not address the issue of coupling *all* of the geometrical objects that describe spacetime to corresponding physical objects, but only examine the role that torsion and energy-momentum play, at least as the theory of teleparallelism, or the translational gauge theory of gravitation, describes the situation. In particular, we shall question the simpler issue of whether it is actually mathematically proper to describe the construction as a true translational gauge theory – i.e., an $R^4$-principal bundle with an $R^4$ connection − as Cho [**2**] attempted to do, or whether one is actually dealing with the "residual" symmetry group of $\{e\}$ that one ultimately reaches in the complement of a singularity set when attempting to define a maximal partial frame field on spacetime. This has a certain intuitive appeal in the eyes of physics, since gravitation is at the same time the weakest of the four fundamental interactions and the one that is left to dictate the structure of spacetime at the astronomical level.

We shall find that the study is necessarily incomplete, since we will be introducing connections on spacetime $G$-structures without discussing how they might be coupled to physical currents, but that is only consistent with the stated focus of this paper. Hopefully, the broader issues that we defer to further research will be better illuminated once we have a clearer description of the role played by translations in the context of manifolds that do not necessarily possess a homogeneous space structure.

A central theme of this article is that although a given manifold $M$ is not generally parallelizable, nevertheless the total space $GL(M)$ of its bundle of linear frames *always* is. Consequently, if one is to avoid the role of topological obstructions in the formulation of teleparallelism then one must essentially "lift" the formalism from the tangent bundle $T(M)$ to the horizontal sub-bundle $H(GL(M))$ of $T(GL(M))$ that one obtains by defining a connection on $GL(M)$. This bundle then represents a sort of "parallelizable covering manifold" for $T(M)$ and the appearance of singularities in maximal partial frame fields on $M$ is then related to rank singularities in the projection of $H(GL(M))$ onto $T(M)$. However, we shall defer a detailed discussion of the connection that was introduced to a later, more comprehensive, study of spacetime structure.

The general outline of this study is to first address the aforementioned issues that were listed at the beginning of the introduction, then to discuss the formalism that one employs when treating spacetime structure in terms of reductions of the bundle of linear frames on the spacetime manifold, or *G-structures*. In particular, we shall discuss variational theory for differential forms on $G$-structures. We will then review how this formalism applies to the theories of gravitation described by general relativity and teleparallelism in their traditional contexts. Finally, we return to the general context of $G$-structures and examine what changes when the parallelizable manifold that we are dealing with is a spacetime $G$-structure $G(M)$, instead of $M$ itself.

**1. Addressing the issues.** The first of the aforementioned four issues can be addressed by asking the simple question of whether it is coordinate systems or local frame fields that play the more fundamental role in physics.

*a. Charts vs. frames.* On closer inspection one notices that quite often the way that one operationally defines a coordinate system in a physics problem is by first defining a frame at a point and "exponentiating" this definition into a coordinate system about this point, which is a process of integration, as one does with a geodesic coordinate system for a connection on the tangent bundle.

In case of the simplest coordinate systems, viz., the Cartesian ones, this integration is essentially invisible since one is dealing with a vector space $V$ to begin with and the coordinate functions are defined by a linear isomorphism of $V$ with $\mathbb{R}^n$. Hence, its differential map is essentially the same linear isomorphism.

However, one also generally makes other simplifying assumptions about the initial frame that force it to produce an integrable situation. In particular, the assumption that a frame is inertial, or "at rest," is such an assumption, since it forces the vector fields of the local frame field that one obtains by parallel translating the chosen frame to the points in its neighborhood to commute; one also calls such a local frame field *holonomic.* However, if one performs an experiment of duration comparable to an Earth day, one will notice unavoidable contributions from the rotation of the Earth, such as Coriolis effects, that belie the assumption that the "laboratory frame" is truly at rest. Indeed, one can see the contradiction to this fact after a few minutes through a common telescope that has not been synchronized for the Earth rotation.

A theoretical physicist should always be aware of the limits of the simplifying assumptions that were made in a mathematical model. There seems to be an "ergodic" principle to the advance of science that says that eventually experimental physics will reach those limits. For instance, one can regard the dawn of quantum theory as the point in time when experimental physics reached the limits of the assumptions that were made in Maxwell's theory of electromagnetism. Perhaps the conjecture that all physical frame fields must have some residual − perhaps one could say "quantum" − degree of non-commutation is justified and that the assumption of perfect commutation is equivalent to the Newtonian assumption that there is such a thing as a universal rest frame.

In the context of modern manifold theory, in which the role of coordinate charts has been "modded out" in the differential structure, the way to resolve the issue of general covariance is to use only those geometrical objects – generally tensor fields of various ranks – that can be defined on the manifold intrinsically, and none that are only definable in the local coordinate charts. The solution to the more complex matter of local frame invariance, when one allows non-integrable − or *holonomic* – local frame fields is to use only those geometrical objects that can be defined by on the bundle $G(M)$ of $G$-frames over the spacetime manifold, where $G$ is some appropriate subgroup of $GL(4)$, and are also $G$-invariant. Examples of such $G$-frames are unit-volume frames for a choice of volume element, orthonormal frames for a choice of metric or pseudo-metric, and even a global frame field, if one exists (for a more detailed discussion of spacetime $G$-structures, see [**3**]).

Since any such reduction of the bundle of linear frames on spacetime to a bundle of $G$-frames – which is called a $G$-structure on $M$ – is a $G$-principal bundle, we are gradually seeing how the geometrical structure of spacetime can be posed as a set of gauge field theories. For a geometrically interesting $G$-structure, there are three fundamental geometrical objects: the fundamental tensor field $t$ that is associated with the reduction, the restriction of the canonical 1-form $\theta^\mu$ on $GL(M)$ to $G(M)$, and the reduced connection $\varpi$ that one obtains by restricting a linear connection $\omega$ on $GL(M)$, if $\omega$ is indeed reducible. When a linear connection is not reducible, it is associated with a "deformation" 1-form $\tau = \omega - \varpi$. What is also quite intriguing is the fact that reductions are associated with effects that amount to spontaneous symmetry breaking, with associated topological defects (cf. [**3**]). A compelling challenge to the gauge theories of spacetime structure is to establish how the geometrical structure might be related to the topological defects as essentially "field sources." One might ponder the way that poking a hole in a plastic material first brings about curvature in the material (or is it really torsion?). Since the shape of the distortion will depend upon mechanical properties of the material, one sees that it is pointless to look for a unique solution of the problem of coupling topological sources to geometrical fields that does not introduce some largely empirical physical element, such as a constitutive law for stress and strain, at least if the solution is to model any physical situation.

*b. Parallelizability.* The issue of non-parallelizability can be resolved for the general spacetime manifold $M$ by observing that regardless of whether the manifold $M$ itself is parallelizable, nevertheless the bundle of linear frames $GL(M)$, as well as its various reductions, is *always* parallelizable; this is equivalent to the fact that $GL(M)$ always admits at least one connection ([1]). Hence, one can use $GL(M)$ as a sort of parallelizable "covering space" for $M$ in such a way that the horizontal sub-bundle $H(GL(M))$ of $T(GL(M))$ that is defined by a choice of linear connection on $GL(M)$ becomes a type of parallelizable "unfolding" of $T(M)$. It is then easy to see how the same information can be contained in the torsion of the connection that makes a global frame field on $GL(M)$ parallel, as well as the torsion and curvature of the connection that defined $H(GL(M))$. In the case of the Levi-Civita connection on the bundle of Lorentz frames, the fact that its torsion is zero, while the "teleparallism" – or *Cartan* – connection has zero curvature but non-zero torsion exhibits the complementarity between the two representations of the same geometrical information: the torsion of one is equivalent to the curvature of the other.

*c (and d). Spacetime translations.* On closer inspection, one sees that because the calculus of variations is sensitive only to *infinitesimal* transformations, it is not actually necessary to define a *global* action of $\mathbb{R}^4$ on the spacetime manifold $M$, but only its infinitesimal generators. This is fortunate because the set of all infinitesimal generators of spacetime translations consists of all vector fields on $M$, which defines a Lie algebra. Although this Lie algebra is infinite-dimensional and quite complicated, it is nevertheless a more tractable construction than the Lie pseudogroup of germs of local diffeomorphisms that one encounters when attempting to integrate the infinitesimal translations into finite ones. Indeed, this integration generally does not produce a global action of $R^4$ on $M$ anyway, although one sometimes requires the action of the aforementioned Lie pseudogroup to be transitive, nonetheless. (In such an event, given any two points $x, y \in M$ there would have to be neighborhoods $U$ of $x$ and $V$ of $y$ and a diffeomorphism $\phi: U \to V$ that takes $x$ to $y$.)

---

[1] One can prove this by a simple partition-of-unity construction (cf. Sternberg [**4**].) However, there is nothing canonical about the resulting connection, so its physical utility is limited beyond the proof of its existence.

If one wishes, as is common in translational gauge theory, to replace a true local action of $R^4$ on the values of the coordinate functions for a chart $(U, x^\mu)$:

$$(\varepsilon^\mu, x^\mu) \infty \ x^\mu + \varepsilon^\mu, \tag{1.1}$$

with an action that depends upon the points of $U$, as is customary in gauge field theories:

$$(\varepsilon^\mu, x^\mu) \infty \ x^\mu + \varepsilon^\mu(x), \tag{1.2}$$

then one should notice that in order to make this last statement mathematically precise, one must realize that it is no longer the group $R^4$ that is acting on the coordinates, but the group $T(U; R^4)$ of smooth maps $\varepsilon: U \infty \ R^4$, under addition of values. We shall call this the *group of local translational gauge transformations.* As opposed to $R^4$, this group, which also defines a vector space, is no longer finite-dimensional; however, it is still Abelian. Since we assumed smoothness, its Lie algebra is also the same vector space of functions on $U$.

When one defines local vector fields on $U$ by way of:

$$\mathbf{v} = \varepsilon^\mu \partial_\mu, \qquad \mathbf{w} = \varepsilon'^\mu \partial_\mu, \tag{1.3}$$

then one sees that although these vector fields commute in the case of constant functions $\varepsilon$ and $\varepsilon'$, in which case, the association of elements of $R^4$ with vector fields on $U$ is a homomorphism, nevertheless, when $\varepsilon$ and $\varepsilon'$ are non-constant, one has:

$$[\mathbf{v}, \mathbf{w}] = (\varepsilon^\mu \partial_\mu \varepsilon'^\nu - \varepsilon'^\mu \partial_\mu \varepsilon^\nu)\partial_\nu, \tag{1.4}$$

which does not vanish unless:

$$\varepsilon^\mu \frac{\partial \varepsilon'^\nu}{\partial x^\mu} = \varepsilon'^\mu \frac{\partial \varepsilon^\nu}{\partial x^\mu}, \tag{1.5}$$

which integrates to:

$$\varepsilon'^\nu - \varepsilon^\nu = \text{constant}. \tag{1.6}$$

This implies that either both functions are constant to begin with or, if they are not, that one is obtained by simply performing a constant shift in values of the other. Hence, in the general case, the map that takes an element of the Lie algebra $T(U; R^4)$ to a vector field in $X(U)$ is not a homomorphism, so the map (1.2) does not define a true infinitesimal action when one interprets the function $\varepsilon^\mu(x)$ as being the components of a vector field on $U$.

**2. Variational formulation of energy-momentum.**  In light of the foregoing discussion, one should ask the more physically fundamental question: "what property of a physical field $\phi$ on $M$ are we trying to describe by saying that it is invariant under spacetime translations?"

In the Newtonian way of looking at motion the conservation of momentum comes about when there is no external force acting on a system.  Hence, if the external force is conservative, its potential function would have to be constant in space and the Lagrangian would depend only upon the velocities of the system components and internal force potentials that could not depend on the absolute positions, but only the relative positions of these components.

Hence, we see that the answer to our rhetorical question is: "the field action $S[\phi]$ that one associates with $\phi$ should not depend upon the points of $M$ *directly*."  Hence, instead of dealing with a Lagrangian of the form $+(x^\mu, \phi, \phi_{,\mu})$, we wish to deal with a Lagrangian of the "vertical $(^2)$" form $+(\phi, \phi_{,\mu})$.

To be more precise in our terminology, we should say that the spacetime transformations that we are dealing with are not true translations, but *convections*, i.e., the local flows that are associated with the vector fields on $M$.  Hence, to say that some geometrical object – say a tensor field $\tau$ – is invariant under convection by a vector field $\mathbf{v}$ is to say that its Lie derivative with respect to $\mathbf{v}$, which we denote by $L_\mathbf{v}\tau$, vanishes.  For example:

$$L_\mathbf{v}f = \mathbf{v}f, \qquad \text{when } f \text{ is a smooth function on } M, \qquad (2.1a)$$
$$L_\mathbf{v}\mathbf{w} = [\mathbf{v}, \mathbf{w}], \qquad \text{when } \mathbf{w} \text{ is a vector field on } M, \qquad (2.1b)$$
$$L_\mathbf{v}\alpha = i_\mathbf{v}d\alpha + di_\mathbf{v}\alpha, \qquad \text{when } \alpha \text{ is a } k\text{-form on } M. \qquad (2.1c)$$

The definition of the Lie derivative extends to higher-rank tensor products by way of:
$$L_\mathbf{v}(\tau \otimes \sigma) = (L_\mathbf{v}\tau) \otimes \sigma + \tau \otimes (L_\mathbf{v}\sigma). \qquad (2.2)$$

For the case at hand, the geometrical object that we are addressing is a 4-form $+$ on the bundle of 1-jets of sections $\phi$ of some bundle $B \infty M$ over spacetime, and we denote this jet bundle by $J^1(M, B)$.  Locally, this bundle has charts that look like $(x^\mu, \psi, p_\mu)$, and for a section $\phi: M \to B$, they look like $(x^\mu, \phi, \phi_{,\mu})$.  One calls the set of all elements in $J^1(M, B)$ that are defined by $\phi$ in this manner the *1-jet prolongation of* $\phi$, and one denotes it by $j^1\phi$.  However, not all of the elements of $J^1(M, B)$ are obtained in this manner.

The *action functional* $S[\phi]$ that is defined by $\phi$, $+$, and a choice of four-dimensional submanifold $N \subsetneq M$ $(^3)$ is then the linear map:

---

$^2$ The word "vertical" is used here in the sense of the bundle of 1-jets of sections of whatever bundle $\phi$ is defined by.

$^3$ A useful generalization in continuum mechanics is to use a differential singular cubic 4-chain for $N$, since one often defines integration on manifolds in such terms at its elementary level, and any compact submanifold can be triangulated by such a chain, at least up to homotopy.

$$S: \Gamma(M; B) \to \mathrm{R}, \quad \phi \infty \quad S[\phi] = \int_N +(j^1\phi). \tag{2.3}$$

In order to make sense of the demand that $S$, or at least $+$, be independent of the points of $M$, at least directly, we say that this means that the Lie derivative of $+$ with respect to *any* vector field $\mathbf{v}$ on $M$ must vanish. Of course, we also must first lift $\mathbf{v}$ to a vector field $\hat{\mathbf{v}}$ on $J^1(M, B)$ that one calls the *1-jet prolongation* of $\mathbf{v}$.

One does this by differentiation, but first one needs to specify how the vector fields on $M$ act on the values of the sections $\phi$, which we now denote by $\phi^k$, $k = 1, \ldots, N$ in order to make the fiber coordinates explicit. Traditionally, one assumes that $\mathbf{v}$ acts trivially on the values of $\phi^k$, but non-trivially on the values of $\phi_{,\mu}$:

$$\delta x^\mu = \varepsilon^\mu(x) \qquad \delta\phi^k = 0, \qquad \delta\phi_{,\mu}^{\ k} = D\phi^k(\delta x^\mu) = \varepsilon^\mu \phi_{,\mu}^{\ k}. \tag{2.4}$$

Hence, we can also represent the 1-jet prolongation of $\mathbf{v}$ to $J^1(M, B)$ in the local form:

$$\hat{\mathbf{v}} = \varepsilon^\mu(x)\left\{ \frac{\partial}{\partial x^\mu} + \phi_{,\mu}^{\ k}\frac{\partial}{\partial p_\mu^k} \right\}. \tag{2.5}$$

Before we examine the consequences of assuming that the 4-form $+$ is invariant under all infinitesimal convections defined by prolongations of vector fields on $M$, we first introduce the *total variation* of $+$ under a variation of $x$ and $\phi^k$ that we represent by a vector field on $B$:

$$\mathbf{v} = \delta x^\mu \frac{\partial}{\partial x^\mu} + \delta\phi^k \frac{\partial}{\partial \phi^k} \tag{2.6}$$

whose 1-jet prolongation is:

$$\hat{\mathbf{v}} = \delta x^\mu \frac{\partial}{\partial x^\mu} + \delta\phi^k \frac{\partial}{\partial \phi^k} + \delta\phi_{,\mu}^{\ k}\frac{\partial}{\partial p_\mu^k}. \tag{2.7}$$

However, in order to be consistent, we must have:

$$\delta(D\phi^k) = D(\delta\phi^k), \tag{2.8}$$

i.e.:

$$\delta\phi_{,\mu}^{\ k} = \partial_\mu \delta\phi^k. \tag{2.9}$$

More specifically, let us assume that $B = \Lambda^p(M, \mathrm{R}^N)$, i.e., $\phi^k$ is a $p$-form on $M$ with values in $\mathrm{R}^N$. If $+ = L(x^\mu, \phi^k, \phi_{,\mu}^{\ k})\mathfrak{I}$, where $L$ is a smooth function on $J^1(\Lambda^p(M, \mathrm{R}^N))$, $\mathfrak{I}$ is a volume element on $M$, that we have pulled up a volume element on $J^1(\Lambda^p(M, \mathrm{R}^N))$ by the projection of that bundle on $M$ then the total variation of $+$ under the infinitesimal action of $\hat{\mathbf{v}}$ is the Lie derivative:

$$\delta+ \quad = \mathrm{L}_{\hat{\mathbf{v}}}\, + \quad = i_{\hat{\mathbf{v}}} d + + di_{\hat{\mathbf{v}}}\, +$$

$$= \mathrm{L}_{\delta x}\, + \; + \; i_{\hat{\mathbf{v}}}\left( \frac{\partial +}{\partial \phi^k} \wedge d\phi^k + \frac{\partial +}{\partial (D\phi^k)} \wedge d(D\phi^k) \right)$$

$$= \mathrm{L}_{\delta x} + \frac{\partial +}{\partial \phi^k} \wedge \delta \phi^k + \frac{\partial +}{\partial (D\phi^k)} \wedge \delta(D\phi^k)$$

$$= \mathrm{L}_{\delta x} + \frac{\partial +}{\partial \phi^k} \wedge \delta \phi^k + \frac{\partial +}{\partial (d\phi^k)} \wedge d(\delta \phi^k)$$

$$= \frac{\partial +}{\partial x^\mu} \delta x^\mu + \frac{\delta +}{\delta \phi^k} \wedge \delta \phi^k + d\left( i_{\delta x^\mu} + (-1)^{4-p-1} \frac{\partial +}{\partial (d\phi^k)} \wedge \delta \phi^k \right). \qquad (2.10)$$

In these computations, we have set $\delta \phi^k = i_{\delta \phi} d\phi^k$, $\delta(D\phi^k) = i_{\delta \phi} d(D\phi^k)$.

Since the last term would revert to a 3-form if the region of integration in $M$ had a boundary, we should say something about the role of boundaries.

In order to derive the field equations for $\phi^k$, we use a vertical variation, i.e., one for which $\delta x^\mu = 0$, and assume that $N$ has no boundary; indeed, one would generally choose $N = M$, or its one-point compactification. If the total variation of $+$ is zero for all such variations of $\phi^k$ then this implies the usual Euler-Lagrange equations for $\phi^k$:

$$0 = \frac{\delta +}{\delta \phi^k} = \frac{\partial +}{\partial \phi^k} - (-1)^{4-p-1} d\left( \frac{\partial +}{\partial (d\phi^k)} \right). \qquad (2.11)$$

In order to obtain the energy-momentum current and canonical stress-energy-momentum tensor field that are associated with $+$ and the infinitesimal action of X($M$) on $M$, we need consider a region $N$ of $M$ that *does* have a boundary and consider non-zero variations of $x^\mu$, as well. If $\phi^k$ is an extremal section, i.e., one that satisfies (2.11), then the total variation of $+$ for an arbitrary variation of $\phi^k$ becomes:

$$\delta + = i_{\delta x} d + + d\left( i_{\delta x} + + (-1)^{4-p-1} \frac{\partial +}{\partial (d\phi^k)} \wedge \delta \phi^k \right) = \frac{\partial +}{\partial x^\mu} \delta x^\mu + d*\mathbf{J} \qquad (2.12)$$

in which we have introduced the 3-form ([4]):

$$*\mathbf{J} = L i_{\delta x} \mathbf{9} + (-1)^{4-p-1} \frac{\partial +}{\partial (d\phi^k)} \wedge \delta \phi^k. \qquad (2.13)$$

If $\delta x^\mu$ is an *arbitrary* vector field on $N$ and the variation of $\phi^k$ is due to the push-forward of $\delta x^\mu$ by $\phi^k$:

$$\delta \phi^k = \phi^k_* \delta x^\mu = D\phi^k(\delta x^\mu), \qquad (2.14)$$

then we have defined a vector field on $N$:

---

[4] Here, we are using the * notation to refer to the Poincaré duality isomorphism *: X($M$) $\to \Lambda^3(M)$, $X \infty$ $i_X \mathbf{9}$, where $\mathbf{9}$ is a volume element on $M$, not the Hodge duality isomorphism that one obtains by further composing with an isomorphism of X($M$) with $\Lambda^1(M)$, as one would derive from a choice of metric tensor field. We shall try to avoid the introduction of a metric unless unavoidable.

$$\mathbf{J} = L\delta x^{\mu} + * \left\{ (-1)^{4-p-1} \frac{\partial +}{\partial (d\phi^{k})} \wedge i_{\delta x} d\phi^{k} \right\} \qquad (2.15)$$

that we call the *energy-momentum* current associated with $\delta x^{\mu}$. We can write this as:

$$\mathbf{J} = T(\delta x^{\mu}), \qquad T = L + * \left\{ (-1)^{4-p-1} \frac{\partial +}{\partial (d\phi^{k})} \wedge d\phi^{k} \right\} \qquad (2.16)$$

in which we have introduced the *canonical stress-energy-momentum tensor $T$* that is associated with $\phi$ and $\mathcal{I}$. If one interprets $\delta x^{\mu}$ as the velocity vector field of some extended object $N$ in $M$ then $T(\delta x^{\mu})$ represents the energy-momentum vector field of the object.

In order to exhibit the actual conservation law that is associated with the invariance of $S[\phi]$ under the infinitesimal action of any $\delta x^{\mu}$, we assume that $\partial N$ is composed of the disjoint union of two diffeomorphic components, $\partial N = B_2 - B_1$, one of which $B_1$ is given the opposite of the induced orientation, that are connected by the flow of $\delta x^{\mu}$ [5]. In order to eliminate the first term in $\delta+$, we must assume that $+$ is independent of $x^{\mu}$ (otherwise, we would have to restrict the class of $\delta x^{\mu}$ to the ones that were "transversal" to $\partial +/\partial x^{\mu}$).

If the integral of the 3-form $*\mathbf{J}$ when taken over $\partial N$ is zero when $*\mathbf{J}$ is defined as in (2.13) and $\phi$ is extremal then its integral over each boundary component is the same:

$$\int_{B_1} *\mathbf{J} = \int_{B_1} *\mathbf{J}. \qquad (2.17)$$

Hence, $*\mathbf{J}$ defines an *absolute integral invariant* of the flow of $\delta x^{\mu}$ that one often calls the *Poincaré-Cartan form*. By Stokes's theorem, this integral over $\partial N$ becomes the action integral and $*\mathbf{J}$ goes to $d*\mathbf{J}$. Hence, when $+$ is independent of $x^{\mu}$ the invariance of the action integral over $N$ under the infinitesimal action of any $\delta x^{\mu}$ is equivalent to the vanishing of $d*\mathbf{J}$, which is equivalent to the vanishing of the divergence of $\mathbf{J}$:

$$\delta\mathbf{J} = *^{-1}d*\mathbf{J}. \qquad (2.18)$$

Hence, the invariance of the action under $\delta x^{\mu}$ is associated with a conserved current that one identifies as the energy-momentum current of the motion.

So far, we have said nothing about the existence of internal − i.e., gauge − symmetries in our field theories and a gauge structure. In the context of spacetime structure, these would come about in the form of the structure group $G$ of the bundle $B$ and its associated $G$-principal fiber bundle $P$, respectively. Since the structure groups that we shall be concerned with will be subgroups of $GL(4)$ and the principal fiber bundles will be reductions of the bundle $GL(M)$ of linear frames on the spacetime manifold, before we recast the previous discussion in context of $G$-structures, we first examine two of the most basic ways that one can approach the geometry of $GL(M)$.

---

[5] One calls such a scenario an elementary example of *Lorentz cobordism,* although more generally one does not assume that the boundary components are diffeomorphic. This has the effect that one must introduce flows with singularities in order to get around the lack of diffeomorphism in some cases.

**3. Geometry on $GL(M)$.** There is a sort of complementarity between the geometry that one obtains by defining a linear connection on $GL(M)$ and the geometry that one obtains by regarding $GL(M)$ as a parallelizable manifold in its own right. In order to exhibit this complementarity explicitly, we first need to put both the geometry of $GL(M)$ when given a linear connection and its geometry as a parallelizable manifold into the same language and then show how these two pictures relate to each other. For the sake of application to the spacetime manifold, we restrict the generality of the discussion to four dimensions, even if most of the statements are more general in scope mathematically.

*a. Linear connection on $GL(M)$.* If we define a linear connection on $GL(M)$ as a $GL(4)$-invariant horizontal complement $H(GL(M))$ to the vertical sub-bundle $V(GL(M))$ of $T(GL(M))$:

$$T(GL(M)) = H(GL(M)) \oplus V(GL(M)) \tag{3.1}$$

then this choice is equivalent to an $\text{Ad}^{-1}$-equivariant 1-form $\omega$ on $T(GL(M))$ that takes its values in the Lie algebra $\mathfrak{gl}(4)$, which we represent by $4 \times 4$ real matrices. Hence, the notation $\omega_\nu^\mu$ will be used, as well. The relationship between $\omega$ and $H(GL(M))$ is defined by setting $H(GL(M))$ equal to the bundle of annihilating subspaces for $\omega$; i.e., the restriction of $\omega$ to any horizontal subspace $H_\mathbf{e}(GL(M))$ must be zero. $\omega$ is also assumed to have the property that its restriction to each vertical subspace of $V(GL(M))$ is a linear isomorphism with $\mathfrak{gl}(4)$. The Cartan structure equations for the torsion 2-form $\Theta^\mu$ and curvature 2-form $\Omega_\nu^\mu$ of $\omega$ are then:

$$\Theta^\mu = \nabla \theta^\mu = d\theta^\mu + \omega_\nu^\mu \wedge \theta^\nu \tag{3.2a}$$

$$\Omega_\nu^\mu = \nabla \omega_\nu^\mu = d\omega_\nu^\mu + \omega_\lambda^\mu \wedge \omega_\nu^\lambda. \tag{3.2b}$$

Here, we have introduced the canonical – or *soldering* – 1-form $\theta^\mu$ on $GL(M)$, which takes a tangent vector $\mathbf{v} \in T_\mathbf{e}(GL(M))$ to the components of its projection $\pi_* \mathbf{v}$ into the tangent space $T_x(M)$ with respect to the frame $\mathbf{e}_x$ in that space; i.e.:

$$\theta^\mu(\mathbf{v})\mathbf{e}_\mu = \pi_* \mathbf{v}. \tag{3.3}$$

When one makes a choice of local section $\mathbf{e}: U \to GL(M)$ of the principal fibration $GL(M) \to M$ over an open subset $U \subset M$ – i.e., a choice of local frame field over $U$ – the 1-form $\theta^\mu$ pulls down to the local coframe field $\theta^\mu$ that is reciprocal to $\mathbf{e}_\mu$:

$$\theta^\mu(\mathbf{e}_\nu) = \delta_\nu^\mu. \tag{3.4}$$

(This is why we use the same notation for both the canonical 1-form on $GL(M)$ and the reciprocal coframe field to a choice of local frame field; hopefully, no confusion will arise.)

The Bianchi identities for $\omega$ then become:

$$\nabla \Theta^\mu = \Omega^\mu_\nu \wedge \theta^\nu \qquad (3.5a)$$

$$\nabla \Omega^\mu_\nu = 0. \qquad (3.5b)$$

*b. Teleparallelism connection on GL(M).*  As mentioned above, the manifold $GL(M)$ is always parallelizable; hence, its own bundle of linear frames $GL(GL(M))$ is trivializable:

$$GL(GL(M)) = GL(M) \mathfrak{Z} (\mathrm{R}^4 \oplus \mathrm{gl}(4)). \qquad (3.6)$$

One can regard the second factor $\mathrm{R}^4 \oplus \mathrm{gl}(4)$ as the underlying vector space for the Lie algebra $\mathrm{a}(4)$ of infinitesimal affine transformations of $\mathrm{R}^4$.  Hence, a choice of global frame field $E_i$, $i = 1, \ldots, 4^2 + 4 = 20$, on $GL(M)$ is equivalent to a 1-form $(^6)$ $\varpi^i$ on $GL(M)$ that takes its values in $\mathrm{a}(4)$ and whose restriction to each tangent space to $GL(M)$ is a linear isomorphism of that tangent space with $\mathrm{a}(4)$.  The 1-form $\varpi^i$ also represents the coframe field that is reciprocal to the frame field $E_i$:

$$\varpi^i (E_i) = \delta^i_j. \qquad (3.7)$$

If we demand that the 1-form $\varpi^i$ be $\mathrm{Ad}^{-1}$-equivariant under the right action of $GL(4)$ on $GL(M)$ then that fact, along with the fact that $\varpi^i$ defines a parallelization of $GL(M)$ makes $\varpi^i$ a *Cartan connection* [**5**].  The difference 1-form:

$$\omega = \varpi - \theta \qquad (3.8)$$

then defines a linear connection on $GL(M)$.  (In order to make sense of the foregoing expression, one must regard $\mathrm{gl}(4)$ and $\mathrm{R}^4$ as subalgebras of $\mathrm{a}(4)$).  The frame field $E_i$ then consists of the *basic* vector fields $E_\mu$, $\mu = 0, \ldots, 3$ on $GL(M)$, relative to a choice of basis for $\mathrm{R}^4$, along with the *fundamental* vector fields $\tilde{E}^\mu_\nu$ that are obtained by choosing a basis for $\mathrm{gl}(4)$.  Individually, these frame fields also trivialize the sub-bundles $H(GL(M))$ and $V(GL(M))$, where we are defining the former by way of $\omega$.  The reciprocal coframe fields for $E_\mu$ and $\tilde{E}^\mu_\nu$ are then $\theta^\mu$ and $\omega^\mu_\nu$, respectively.

Let us proceed with the usual definitions for a connection defined by a global framefield.  To be precise, the connection that we are defining is defined on $GL(GL(M))$, but since that bundle is trivial, we deal with all of the geometrical expressions after they have been pulled down to $GL(M)$ by way of the frame field $E_i$.

The connection 1-form $\omega_\parallel$ that we introduce is then a 1-form on $GL(M)$ with values in $\mathrm{a}(4)$ that satisfies $(^7)$:

$$DE_i = -[\omega_\parallel]^j_i \otimes E_j. \qquad (3.9)$$

---

The matrix notation for $\omega_{\parallel}$ refers to the representation of $\mathfrak{a}(4)$ as $5 \times 5$ real matrices of the form:

$$[\omega_{\parallel}] = \begin{bmatrix} \omega^{\mu}_{\nu} & \theta^{\mu} \\ 0 & 1 \end{bmatrix} \tag{3.10}$$

and the action of this matrix on $E_i$ is:

$$[\omega_{\parallel}]_i^{\,j} \otimes E_j = \begin{cases} \omega^{\nu}_{\mu} \otimes E_{\nu} & i = 1,2,3,4 \\ \omega^{\nu}_{\mu} \otimes E^{\mu}_{\nu} & i > 4. \end{cases} \tag{3.11}$$

The structure equations for $\omega_{\parallel}$ then give:

$$\Theta^i_{\parallel} = d\varpi^i \tag{3.12a}$$

$$\Omega_{\parallel} = 0, \tag{3.12b}$$

and the corresponding Bianchi identities are:

$$\nabla \Theta^i_{\parallel} = 0 \qquad\qquad \nabla \Omega_{\parallel} = 0. \tag{3.13}$$

In order to relate the structure equations for $\omega_{\parallel}$ to the structure equations for $\omega$, one need only note that if $\varpi = \omega + \theta$ then:

$$\Theta_{\parallel} = d\varpi = d\omega + d\theta = (\Omega - \omega \wedge \omega) + (\Theta - \omega \wedge \theta), \tag{3.14}$$

or, in matrix form:

$$\Theta_{\parallel} = \begin{bmatrix} \Omega^{\mu}_{\nu} & \Theta^{\mu} \\ 0 & 0 \end{bmatrix} - \begin{bmatrix} \omega^{\mu}_{\lambda} \wedge \omega^{\lambda}_{\nu} & \omega^{\mu}_{\nu} \wedge \theta^{\nu} \\ 0 & 0 \end{bmatrix}. \tag{3.15}$$

Hence, the torsion 2-form of $\varpi$ contains information from both the torsion and curvature 2-forms of $\omega$. Then again, one must keep in mind that the dimension of the tangent spaces to $GL(M)$ equals the sum of the dimensions to the tangent spaces to $M$ and the Lie algebra $\mathfrak{gl}(4)$, so all we are doing, in effect, is concatenating the information in $\omega$ and $\theta$ into the information in $\omega_{\parallel}$. Furthermore, since $\omega$ vanishes on horizontal vectors, the restriction of $\Theta_{\parallel}$ to $H(GL(M))$ is simply:

$$\Theta_{\parallel} = \begin{bmatrix} \Omega^{\mu}_{\nu} & \Theta^{\mu} \\ 0 & 0 \end{bmatrix}. \tag{3.16}$$

We now address the difference between defining a global frame field on $GL(M)$, as we have, and defining a global frame field on $M$, as is customary in teleparallel theories of gravitation. By the local triviality of $GL(M)$, one can always find an open neighborhood $U$ of any point $x \in M$ over which a local frame field $\mathbf{e}: U \to GL(M)$ exists. One considers all such open neighborhoods for a given point as partially ordered by inclusion and then partially orders all of the local sections over these subsets that agree by restriction ([8]).  If one attempts to characterize the maximal neighborhoods of this partial ordering, one finds that ultimately one arrives at open subsets whose complements are closed subsets of $M$ that represent singular subsets for the partial frame field that one is trying to maximize.  They can also be triangulated into *singularity complexes* for the partial frame field, which are closed and homologous cochains in $Z_2$-homology; hence, they define a $Z_2$–homology class of mixed type, in general.  By Poincaré-Alexander duality, they are dual to the total Stiefel-Whitney class in $Z_2$-cohomology.  In the simplest scenario, such as partial frame fields on $S^{2n}$, the singularity complex consists of a finite set of isolated points that is dual to the Euler class, as in the Poincaré-Hopf theorem.

We could regard $T(M)$ as having been replaced by $H(GL(M))$, in which we are using the fibers of $GL(M)$ to unfold the non-parallelizable vector bundle $T(M)$ into a parallelizable one.  The restriction of the canonical 1-form $\theta^\mu$ to $H(GL(M))$ then replaces a choice of coframe field on $M$, as in the usual teleparallelism constructions.  However, the main difference is that unless $M$ is parallelizable to begin with, the topological information that a singularity complex represents shows up as the geometrical information that the differential, *a fortiori*, the exterior derivative of $\theta^\mu$ includes a contribution from the fiber dimension, in form of curvature, as one sees in (3.16).

Ultimately, one must ask the question of whether a connection on $GL(M)$ with zero curvature can actually exist to begin with, or does topology obstruct such a possibility.  For instance, the Gauss-Bonnet-Chern theorem precludes the existence of such a connection on any compact orientable even-dimensional manifold, since its Euler-Poincaré characteristic will be non-vanishing.

One can also regard curvature as an integrability condition for the horizontal sub-bundle.  If one looks at the commutation relations for the basic vector fields $E_\mu$ that trivialize $H(GL(M))$:

$$[E_\mu, E_\nu] = -\Theta^\lambda(E_\mu, E_\nu)E_\lambda - \Omega_\sigma^\rho(E_\mu, E_\nu)\tilde{E}_\rho^\sigma, \qquad (3.17)$$

then one sees that torsion obstructs the basic frame field from being Abelian and curvature obstructs it from being involutive; by contrast, for a frame field on $M$, only the first term in the right-hand side would appear.  Note that one can regard the vector fields $E_\mu$ as the infinitesimal generators of parallel translations of the frames on $M$ and equation (3.17) then defines the Lie subalgebra of the Lie algebra of vector fields on $GL(M)$ that it generates.

*c.  Introduction of a metric.*  Since Einstein's theory of gravitation centers around the importance of the spacetime metric tensor field, we need to reconcile this fact with the foregoing pre-metric discussion.  We shall use the language of $G$-structures [**4, 6, 7**], i.e., reductions of $GL(M)$ to sub-bundles whose structure group is a subgroup $G$ in $GL(4)$.

Really, one should regard $GL(M)$ as a reduction of the bundle $A(M)$ of affine frames on $M$ to begin with.  However, the reduction is topologically unobstructed and unique, and the geometrical information contained in a connection $\varpi$ on $A(M)$ is equivalent to the information that is contained in $\theta^\mu$ and $\omega = \overline{\omega} - \theta$ on $GL(M)$. Hence, the only significant difference between the two bundles is in the fact that the action of $\mathbf{R}^4$ on $A(M)$ is more natural than its action on $GL(M)$, viz., one can translate the *origin* of an affine frame $\mathbf{e}_\mu$ by the tangent vector $\mathbf{a} = a^\mu \mathbf{e}_\mu$, but the action of $a^\mu$ on $\mathbf{e}_\mu$ itself is less clear. In physics, this fact is quite relevant to all discussions of energy-momentum, which is presumed to relate to the action of $\mathbf{R}^4$. Hence, we now see that in order to apply Nöther's theorem to $\mathbf{R}^4$ directly, we must define our source-field action on $A(M)$ in order for the action of $\mathbf{R}^4$ to be meaningful except in the case of highly homogeneous spacetime manifolds. However, as we have observed, it seems to be the infinitesimal convections that play the fundamental role on $M$, not $\mathbf{R}^4$, so we make the first reduction from $A(M)$ to $GL(M)$, anyway.

---

[8] What we are defining is something like a presheaf of local sections of $GL(M)$, but instead of passing to the local limit – viz., germs of local sections – we are passing to the opposite limit.

Before we further reduce $GL(M)$ to a bundle of Lorentz frames, we first reduce it to a bundle of oriented linear frames, $GL^+(M)$, by choosing a orientation on $GL(M)$, and a bundle of unit-volume frames, $SL(M)$, by choosing a volume element on $GL^+(M)$.

For the first reduction, we define a $GL(4)$-equivariant map:

$$\text{sign}: GL(M) \rightarrow GL(4)/GL^+(4) = \mathbf{Z}_2 = \{+, -\}, \tag{3.18}$$

and then define the reduced bundle $GL^+(M)$ as $\text{sign}^{-1}(+)$.

At this point, we pause to observe that reductions do not always exist, and are generally topologically obstructed by Stiefel-Whitney classes. In the present case, one cannot reduce to $GL^+(M)$ unless $M$, or rather $T(M)$, is orientable, and for a compact $M$ this is obstructed by the first Stiefel Whitney class of $T(M)$. Furthermore, reductions are not always unique, even up to homotopy. In the present case, since there is nothing special about either component of $GL(M)$, we could have also defined $GL^+(M)$ as $\text{sign}^{-1}(-)$ and obtained an isomorphic principal bundle.

For the second reduction, we define a $GL^+(4)$-equivariant map:

$$\det: GL^+(M) \rightarrow GL^+(4)/SL(4) = \mathbf{R}^+ \tag{3.19}$$

that takes an oriented linear frame to the determinant of its matrix with respect to any other oriented linear frame, and then define $SL(M)$ as $\det^{-1}(1)$. One can associate this map with a *fundamental tensor field* $\mathfrak{V}$ on $M$ by way of:

$$\mathfrak{V} = \sqrt{|\det(\mathbf{e}_\mu)|}\, \theta^0 {}^\wedge \theta^1 {}^\wedge \theta^2 {}^\wedge \theta^3 = \frac{1}{4!}\sqrt{|\det(\mathbf{e}_\mu)|}\, \varepsilon_{\mu\nu\rho\sigma} \theta^{\mu\wedge} \theta^{\nu\wedge} \theta^{\rho\wedge} \theta^\sigma, \tag{3.20}$$

i.e., a volume element on $M$. Clearly, any other choice of element in $\mathbf{R}^+$ would produce an isomorphic reduction, but since $SL(4)$ is a deformation retract of $GL^+(4)$, they are all homotopic. Because of this latter fact, if the reduction to $GL^+(M)$ exists, so does the reduction to $SL(M)$. Nevertheless, one should keep in mind that although the freedom to choose a unit of volume arbitrarily has no topological significance, it has considerable geometrical significance, and represents a type of $\mathbf{R}^+$ gauge in physical field theory.

In order to reduce to a bundle of Lorentzian frames on $M$, one must first define an $SL(4)$-equivariant map:

$$g: SL(M) \rightarrow SL(4)/SO(3,1), \tag{3.21}$$

that takes an oriented unit volume frame $\mathbf{e}_\mu$ to a symmetric non-degenerate matrix $g_{\mu\nu}$ that is similar to $\eta_{\mu\nu} = \mathrm{diag}(+1, -1, -1, -1)$, i.e., the matrix of the Minkowski space scalar product with respect to the canonical frame field on $\mathbf{R}^4$. The reduced bundle of oriented Lorentzian frames, $SO(3,1)(M)$, is then defined to be $g^{-1}(\eta_{\mu\nu})$. To see that this is essentially the gist of the "vierbein" approach to the definition of a spacetime metric, we note that the fundamental tensor field on $M$ that this reduction defines (and is equivalent to) is the Lorentzian metric tensor field:

$$g = g_{\mu\nu}(\mathbf{e}_\mu)\theta^\mu \otimes \theta^\nu, \tag{3.22}$$

and for the frames of $SO(3,1)(M)$, i.e., the oriented Lorentzian frames, this becomes:

$$g = \eta_{\mu\nu}\theta^\mu \otimes \theta^\nu. \tag{3.23}$$

Note that one does not have to actually define a global frame field on $M$ in order to define a metric tensor field. In effect, by the $SL(4)$-equivariance of $g$ one only needs to define an *equivalence class* of frames at each point, namely, the orbit of some frame under the action of the special Lorentz group. This points to the non-uniqueness of the reduction since any frame could just as well be *defined* to be Lorentz-orthonormal and then one simply uses the $SO(3,1)$ orbit that it defines to represent the metric at that point. As far as homotopy equivalence of the choices, note that the homogeneous space $SL(4)/SO(3,1)$, which represents the space of special Lorentzian strains, is not actually contractible, unlike the Riemannian case of $SL(4)/SO(4)$, so the non-vanishing homotopy groups obstruct the uniqueness of the reduction, as well as its existence. For a non-compact orientable manifold there is no obstruction to the existence of a Lorentzian metric, but for a compact orientable manifold, the obstruction of the Euler class − or equivalently, the top Stiefel-Whitney class − of $T(M)$. The homotopy classes are then indexed by $H^3(M; Z_2)$, at least in the case of four dimensions.

It is important to note that the geometrical information at any level of reduction essentially subsumes the information at the higher levels. For instance, given a Lorentzian metric tensor field $g$, one can define a Lorentzian volume element by way of:

$$\boldsymbol{\vartheta} = \sqrt{|\det g|}\,\theta^0 \wedge \theta^1 \wedge \theta^2 \wedge \theta^3 = \frac{1}{4!}\sqrt{|\det g|}\,\varepsilon_{\mu\nu\rho\sigma}\theta^\mu \wedge \theta^\nu \wedge \theta^\rho \wedge \theta^\sigma; \tag{3.24}$$

for an orthonormal coframe field $\theta^\mu$ this reduces to

$$\boldsymbol{\vartheta} = \theta^0 \wedge \theta^1 \wedge \theta^2 \wedge \theta^3. \tag{3.25}$$

One might finally reach the ultimate level of reduction of $GL(M)$, for which the subgroup is $\{e\}$ itself, namely, a global frame field $\mathbf{e}_\mu$ on $M$. This reduction is generally obstructed by all four Stiefel-Whiteney classes of $T(M)$; indeed, their vanishing is necessary, but not sufficient, for the parallelizability of $M$. Furthermore, since any two frame fields are related by an element of $GL(4)$ at every point, the homotopy classes of global frame fields on $M$ are in one-to-one correspondence with the homotopy classes of maps from $M$ to $GL(4)$, which is equivalent to the homotopy classes of maps from $M$ to $SO(4)$.

Given a global frame field $\mathbf{e}_\mu$, one can define a Lorentzian metric tensor field by way of (3.23) – i.e., by declaring the frame field to be Lorentz-orthonormal – and a Lorentzian volume element by way of (3.25).

*d. Reduction of connections.* If one is given a linear connection $\omega$ on $GL(M)$, one can also address the issue of whether and how $\omega$ reduces from a connection on $GL(M)$ to a connection on $G(M)$ for some subgroup $G$ in $GL(4)$. Just as the homogeneous space $GL(4)/G$ plays an important role in the reduction of the bundle, the vector space $\mathrm{gl}(4)/\mathrm{g}$ plays an important role in the reduction of connections. This vector space is linearly isomorphic to the tangent spaces of $GL(4)/G$ and represents essentially "non-isomorphic deformations" of the reduction. Of particular interest, and widespread utility, is the case in which the Lie algebra $\mathrm{g}$ is *reductive* in $\mathrm{gl}(4)$ **[8]**. In this case, there is a complementary vector subspace $\mathrm{m}$ to $\mathrm{g}$ in $\mathrm{gl}(4)$ – so $\mathrm{gl}(4) = \mathrm{g} \oplus \mathrm{m}$ – and the adjoint action of $\mathrm{g}$ on $\mathrm{m}$ obeys $[\mathrm{g}, \mathrm{m}] = \mathrm{m}$. For a reductive subalgebra $\mathrm{g}$, the quotient space $\mathrm{gl}(4)/\mathrm{g}$ is linearly isomorphic to $\mathrm{m}$. Since most of the geometrically interesting reductions involve subgroups that are reductive in the preceding subgroup – for instance, the reductions above, except for the full reduction from $GL(4)$ to $\{e\}$ – we confine our attention to the reductive case.

When $\mathrm{g}$ is a reductive subalgebra of $\mathrm{gl}(4)$, one can decompose and $\omega$ into the sum $\varpi + \tau$ of a 1-form $\varpi$ that takes its values in $\mathrm{g}$ and a 1-form $\tau$ that takes its values in $\mathrm{m}$. If we restrict $\omega$ to the reduced bundle $G(M)$ then a necessary and sufficient condition for $\omega$ to be a $G$-connection on $G(M)$ is that $\omega$ take its values in $\mathrm{g}$; i.e., that $\tau = 0$. When the reduction is associated with a fundamental tensor field $t$, the reducibility of $\omega$ is also equivalent to ([9]):

$$Q = \nabla t = dt + \omega \wedge t = 0. \tag{3.26}$$

For instance, when $t$ is actually $g$ we are saying that a linear connection must have vanishing non-metricity in order for it to be a metric connection. In order to deal only with objects that are defined on $GL(M)$, and not their representatives on $M$, we continue to regard a metric on $M$ as an $SL(4)$-equivariant map $g$ from oriented linear frames $\mathbf{e}_\mu$ to component matrices $g_{\mu\nu}$. If we use the canonical 1-form on $GL(M)$ as a replacement for a global coframe field on $M$ then we can then represent the metric tensor field as an object on $GL(M)$ by way of $g_{\mu\nu}\,\theta^\mu {\otimes} \theta^\nu$, which reduces to $\eta_{\mu\nu}\,\theta^\mu {\otimes} \theta^\nu$ for the bundle of Lorentzian frames. Similarly, the non-metricity $Q$ becomes a 1-form on $GL(M)$ with values in the space of infinitesimal traceless Lorentzian strains, and is defined by:

$$Q_{\mu\nu} = dg_{\mu\nu} - \omega^\lambda_\mu\, g_{\lambda\nu} - \omega^\lambda_\nu\, g_{\lambda\mu}. \tag{3.27}$$

---

[9] Of course, the precise meaning of the exterior product in this expression depends upon the way that the Lie algebra $\mathrm{gl}(4)$ acts upon the vector space in which $t$ takes its values.

As usual there is a unique linear connection $\varpi$ with the properties that $Q = 0$ and $\Theta^\mu = 0$ relative to the reduction to the bundle $SO(3,1)(M)$, namely the Levi-Civita connection. It can be represented as an object on $GL(M)$ as a 1-form $\varpi_\nu^\mu$ with values in so(3,1) that satisfies:

$$g_{\mu\nu} d\theta^\nu = d(g_{\mu\nu}\,\theta^\nu) - dg_{\mu\nu} \wedge \theta^\nu = \varpi_{\mu\nu} \wedge \theta^\nu, \tag{3.28}$$

where we have set $\varpi_{\mu\nu} = g_{\mu\lambda}\varpi_\nu^\lambda$. For a general frame in $SL(M)$, one has, from (3.27):

$$\varpi_{\mu\nu} + \varpi_{\nu\mu} = dg_{\mu\nu}. \tag{3.29}$$

For a Lorentzian frame, this becomes:

$$\varpi_{\mu\nu} + \varpi_{\nu\mu} = 0, \tag{3.30}$$

i.e., $\varpi_{\mu\nu}$ is antisymmetric, which is equivalent to saying that $\varpi_\nu^\mu \in$ so(3,1).

We shall define the coframe field on $M$:

$$e_\mu = g_{\mu\nu}\,\theta^\nu \tag{3.31}$$

to be the *Lorentz-reciprocal* coframe field to $\mathbf{e}_\mu$; for a Lorentzian frame, it differs from the usual (Euclidian) reciprocal coframe field by an orientation; equation (3.31) also defines an $\mathbf{R}^4$-valued 1-form on $SL(M)$. This makes the first of the defining equations, (3.28) and (3.29), for the Levi-Civita connection on $SL(M)$ take the concise form:

$$de_\mu = (dg_{\mu\nu} + \varpi_{\mu\nu}) \wedge \theta^\nu, \tag{3.32}$$

and for Lorentzian frames this becomes:

$$\tilde{\nabla} e_\mu = de_\mu - \varpi_\mu^\nu \wedge e_\nu = 0. \tag{3.33}$$

Any metric connection $\omega$ on $SL(M)$ can be expressed as the Levi-Civita connection $\varpi$ plus a difference 1-form $\tau = \omega - \varpi$ that also takes its values in $SO(3,1)$; this 1-form is often referred to as the *contortion* tensor for $\omega$. It can be expressed in terms of the torsion of $\omega$ by solving the equation:

$$\Theta^\mu = \tau_\nu^\mu \wedge \theta^\nu \tag{3.34}$$

for $\tau$. We set:

$$\Theta^\mu = \tfrac{1}{2} S_{\rho\sigma}^\mu\,\theta^{\rho} \wedge \theta^\sigma, \qquad \tau_\nu^\mu = C_{\nu\sigma}^\mu\,\theta^\sigma, \tag{3.35}$$

and observe that $C_{\nu\sigma}^\mu$ does not have to be antisymmetric in its lower indices as $S_{\nu\sigma}^\mu$ does, so we have:

$$\tau_\sigma^\mu \wedge \theta^\sigma = \tfrac{1}{2}(C_{\rho\sigma}^\mu - C_{\sigma\rho}^\mu)\,\theta^{\rho} \wedge \theta^\sigma. \tag{3.36}$$

The equation:

$$S_{\rho\sigma}^\mu = C_{\rho\sigma}^\mu - C_{\sigma\rho}^\mu \tag{3.37}$$

can then be solved by symmetrization:

$$C_{\rho\sigma}^\mu = \tfrac{1}{2} g^{\mu\lambda}(S_{\lambda\rho\sigma} + S_{\rho\sigma\lambda} + S_{\sigma\lambda\rho}). \tag{3.38}$$

We can then put this into the form:

$$\tau_\nu^\mu = \tfrac{1}{2} g^{\mu\lambda}(S_{\lambda\nu\sigma}\theta^\sigma + S_{\nu\sigma\lambda}\theta^\sigma + S_{\sigma\lambda\nu}\theta^\sigma)$$

$$= \tfrac{1}{2}\{i_\nu\Theta^\mu - g^{\mu\lambda}(g_{\rho\nu}i_\lambda\Theta^\rho + g_{\rho\sigma}\Theta^\rho(\mathbf{e}_\lambda, \mathbf{e}_\nu)\theta^\sigma)\}, \qquad (3.39)$$

or, more concisely:

$$\tau_{\mu\nu} = \tfrac{1}{2}\{i_\nu\Theta_\mu - i_\mu\Theta_\nu - g_{\rho\sigma}\Theta^\rho(E_\mu, E_\nu)\theta^\sigma\}. \qquad (3.40)$$

(In these expressions, we have abbreviated $i_{E_\mu}$ by $i_\mu$.)

If a connection $\omega$ is reducible under a given bundle reduction then it is easy to see that its torsion and curvature reduce by simple restriction, since they are obtained by local operations on $\omega$. In particular, the Riemannian curvature 2-form on $SO(3,1)(M)$ takes the form:

$$\tilde{\Omega}_\nu^\mu = \tfrac{1}{2} R_{\nu\rho\sigma}^\mu \, \theta^\rho \wedge \theta^\sigma. \qquad (3.41)$$

If $\omega$ is an arbitrary metric connection on $SL(M)$ with a difference form $\tau$ relative to the Levi-Civita connection then we can also express its torsion and curvature in terms of the difference form:

$$\Theta^\mu = \tau \wedge \theta^\mu \qquad (3.42a)$$

$$\Omega = \tilde{\Omega} + d\tau + \varpi \wedge \tau + \tau \wedge \varpi + \tau \wedge \tau. \qquad (3.42b)$$

In particular, if $\omega$ is the teleparallelism connection this entails that:

$$\tilde{\Omega} = -(d\tau + \varpi \wedge \tau + \tau \wedge \varpi + \tau \wedge \tau) \qquad (3.43)$$

The *Ricci curvature* that is associated with $\tilde{\Omega}_\nu^\mu$ becomes a 1-form on $SO(3,1)(M)$ with values in $\mathbf{R}^4$; it is defined by:

$$\mathbf{R}_\nu = i_\mu\tilde{\Omega}_\nu^\mu = R_{\nu\rho\mu}^\mu \theta^\sigma = R_{\nu\mu}\theta^\mu. \qquad (3.44)$$

By the $SO(3,1)$-equivariance of $\tilde{\Omega}_\nu^\mu$ this expression does not actually depend upon the choice of basis for $\mathbf{R}^4$ that we used to define the basic vector fields $E_\mu$. As usual, the component matrix $R_{\nu\mu}$ is symmetric.

The *scalar curvature* can then be defined by:

$$R = g^{\mu\nu} i_\mu\mathbf{R}_\nu = R_\nu^\mu. \qquad (3.45)$$

Because this function is independent of the choice of frame in $SO(3,1)(M)$, it also defines a smooth function on $M$.

We can then define the Einstein tensor field on $SL(M)$ as the $\mathbf{R}^4$-valued 1-form:

$$\mathbf{G}_\mu = \mathbf{R}_\mu - \tfrac{1}{2}Re_\mu = (R_{\mu\nu} - \tfrac{1}{2}Rg_{\mu\nu})\theta^\nu. \qquad (3.46)$$

We point out that when we return to the foregoing constructions in the context of a parallelizable $M$ the only change in form of the aforementioned expressions will amount to the replacement of $E_\mu$ with $\mathbf{e}_\mu$, the canonical 1-form $\theta^\mu$ with the reciprocal coframe field to $\mathbf{e}_\mu$, and all of the differential forms defined on $GL(M)$ with their pull-downs to $M$ by way of $\mathbf{e}_\mu$. It is precisely this formal analogy that allows us to isolate the role of the topology of $M$ most concisely in the form of the structure of the singularities of the projection of the basic frame field on $GL(M)$ onto $M$.

**4. Gravitation.** In order to complete the Einstein equations, we need to define the stress-energy-momentum tensor as an $R^4$-valued 1-form on $SL(M)$. This is not too much of a stretch from the usual definition of that tensor field on $M$ as having the form:

$$T = T_{\mu\nu}\theta^\mu \otimes \theta^\nu, \tag{4.1}$$

in which the component matrix $T_{\mu\nu}$ is symmetric. In the right-hand side of (4.1) $\theta^\mu$ refers to a local coframe field on $M$, but by the $SL(M)$-equivariance of $T_{\mu\nu}$ under a change of frame, $T$ is equivalent to an expression that is defined globally on $SL(M)$:

$$T(\mathbf{e}_\lambda) = T_{\mu\nu}(\mathbf{e}_\lambda)\theta^\mu \otimes \theta^\nu, \tag{4.2}$$

which allows us to define an associated $R^4$-valued 1-form:

$$\mathbf{t}_\mu = T_{\mu\nu}(\mathbf{e}_\lambda)\theta^\nu. \tag{4.3}$$

Einstein's equation of gravitation then takes the form:

$$\mathbf{G}_\mu = l^2 \mathbf{t}_\mu, \tag{4.4}$$

in which we have introduced the Planck length $l$ that satisfies $l^2 = 8\pi\kappa$.

The challenge at hand is to relate this equation to a corresponding equation in terms of the teleparallelism connection $\varpi_\parallel$ that is associated with $\varpi$. We ignore the topological obstructions to parallelizability of the spacetime manifold $M$ and assume that either it is parallelizable or that discussion is at best restricted to the domain of a maximal partial section. We then simply review the usual sort of constructions that define teleparallel gravitation as a translation gauge theory of gravity and relate it to Einstein's theory.

Hence, we assume that we have a frame field $\mathbf{e}_\chi$ with a reciprocal coframe field $\theta^\mu$, which defines a Lorentzian metric $g = \eta_{\mu\nu}\theta^\mu \otimes \theta^\nu$ and a Lorentzian volume element $\mathbf{9}$ $= \sqrt{|\det g|}\ \theta^0 \wedge \theta^1 \wedge \theta^2 \wedge \theta^3$. As we observed above, the essential change in form of the geometry of $GL(M)$ that we described in section 3 is the replacement of the basic vector fields $E_\mu$ with the global frame field $\mathbf{e}_\mu$, the canonical 1-form $\theta^\mu$ with the coframe reciprocal to $\mathbf{e}_\mu$, and the differential forms on $GL(M)$ get replaced with their pull-downs to $M$ by way of $\mathbf{e}_\mu$.

*a. Translational gauge.* As we have previously observed, a better word to use than "translational" would be "convectional" since the translations that are being described seems to pertain to the local flows of vector fields on $M$. In any event, we are more concerned with the infinitesimal generators, namely the Lie algebra $X(M)$ of vector fields on $M$.

The frame members $\mathbf{e}_\mu$ generate a subalgebra of $X(M)$ by their Lie brackets, which also defines a 2-form $\Theta^\mu$ on $M$ with values in $\mathbb{R}^4$ by way of:

$$[\mathbf{e}_\mu, \mathbf{e}_\nu] = -\Theta^\mu(\mathbf{e}_\mu, \mathbf{e}_\nu)\mathbf{e}_\lambda, \tag{4.5}$$

as well as a map:

$$c: M \to \mathbb{R}^4 \wedge \mathbb{R}^4 \otimes \mathbb{R}^4, \qquad x \infty \quad c_{\mu\nu}^\lambda(x) = -\Theta^\mu(\mathbf{e}_\mu, \mathbf{e}_\nu) \tag{4.6}$$

that one may call the *structure function* that is associated with the frame field $\mathbf{e}_\mu$. In the event that $M$ is a group manifold and $\mathbf{e}_\mu$ is left-invariant this function would be constant and the value it takes would be the array of structure constants for its Lie algebra relative to that choice of basis. The very fact that we are allowing for the possibility that the right-hand side of (4.5) might be non-zero is precisely why we are not actually using the vector fields $\mathbf{e}_\mu$ as the fundamental vector fields of the action of $\mathbb{R}^4$ on $M$, since such an action would differentiate to a homomorphism of the Lie algebra of $\mathbb{R}^4$ into the Lie algebra $X(M)$.

We use the notation $\Theta^\mu$ under advisement since it coincides with our notation for the torsion 2-form of a connection. However, the geometry of a parallelizable manifold can be described in terms of a connection on its linear frame bundle that makes a choice of global frame field parallel. Since that bundle is trivial, it is more concise to deal with the geometrical objects that are pulled down to $M$ by way of the section. In particular, the linear connection is a 1-form $\omega$ on $M$ that takes its values in $\mathfrak{gl}(4)$ and satisfies the defining constraints:

$$D\mathbf{e}_\mu = \omega_\mu^\nu \otimes \mathbf{e}_\nu, \tag{4.7a}$$

$$D\theta^\nu = -\omega_\mu^\nu \otimes \theta^\mu. \tag{4.7b}$$

The structure equations of this connection then give:

$$\Theta^\mu = d\theta^\mu, \qquad \Omega = 0. \tag{4.8}$$

Actually, the first equation also follows as a natural consequence of the properties of the exterior derivative operator:

$$d\theta^\lambda(\mathbf{e}_\mu, \mathbf{e}_\nu) = \mathbf{e}_\mu\theta^\lambda(\mathbf{e}_\nu) - \mathbf{e}_\nu\theta^\lambda(\mathbf{e}_\mu) - \theta^\lambda([\mathbf{e}_\mu, \mathbf{e}_\nu]) = -\theta^\lambda([\mathbf{e}_\mu, \mathbf{e}_\nu]) = \Theta^\lambda(\mathbf{e}_\mu, \mathbf{e}_\nu). \tag{4.9}$$

One sees that the essential information in this type of geometry is contained in the torsion 2-form, which is deeply rooted in the structure of the Lie algebra of vector fields on $M$. The sense in which we can call a parallelizable manifold an "almost Lie group" is that we are using parallel translation as a substitute for left-translation, parallel vector fields as substitutes for left-invariant vector fields, and the structure function as a substitute for the structure constants. Indeed, a compact parallelizable manifold admits a frame field whose structure function is constant iff it is diffeomorphic to a compact Lie group (cf. [**10**]).

*b. Lagrangian considerations.* In order to make contact with gauge field theories, if we are assuming that our ultimate field Lagrangian (i.e., one that accounts for both spacetime structure and source distributions) is invariant under all convections then we might say that we are introducing $\theta^\mu$ as the "gauge potential" that must be introduced in order to insure this invariance and $\Theta^\mu$ as its "field strength" 2-form. Interestingly, the connection $\omega$ that we introduced seems to play no further role, except insofar as it justifies calling $\Theta^\mu$ a torsion 2-form.

Here, we pause to ponder the subtlety that arises in the last statement. First, we briefly summarize the formalism of gauge field theories in their general sense.

In conventional field theories, one has a Lie group $G$ of internal field symmetries – i.e., transformations of the space $V$ in which the fields take their its values – and a $G$-principal bundle $P \to M$ over some configuration manifold $M$; one refers to bundle as the *gauge structure* for the field theory. In order to maintain the invariance of the field Lagrangian under the action of the gauge group on $P$ and $V$, one must introduce a g-valued connection 1-form $\omega$ on $P$ and replace all derivatives with covariant derivatives using $\omega$ ("minimal coupling"). One interprets the connection 1-form as a *gauge potential*, and its curvature $\Omega$ as the *gauge field strength* that is associated with it. This geometrical object is assumed to be the physically meaningful field because it is invariant under gauge transformations of $\omega$, just as the electric field strength vector field is more physically fundamental than a choice of electric potential function.

In order to make this geometrical model into a physical model, one must introduce some mechanism for the generation of the field $\Omega$, i.e., a source term. Customarily, the source term is a 1-form $J$ with values in g, and the field equation, which is due to Yang and Mills, by way of analogy with electromagnetism, is:

$$*\nabla*\Omega = J. \tag{4.10}$$

The second of the Maxwell equations, $d\Omega = 0$, which generalizes to $\nabla\Omega = 0$, is actually an identity in the geometrical formulation, namely, the Bianchi identity for curvature.

In the present case, it has been assumed that $G = \mathbf{R}^4$, even though we have already observed that we are actually dealing with the Lie pseudogroup of germs of local flows of either vector fields on $M$ or horizontal vector fields on $GL(M)$ relative to a choice of linear connection. Nevertheless, we actually use the $SO(3,1)$-principal bundle $SO(3,1)(M)$ as our gauge structure when one has chosen an orientation and a Lorentzian metric on $T(M)$; this suggests that the true "gauge group" is $SO(3,1)$. The appearance of the 1-form $\theta^\mu$ is actually canonical, in that it is defined by the structure of the bundle $GL(M)$ itself, not a consequence of any invariance under the action of $\mathbf{R}^4$. Indeed, the canonical 1-form on $GL(M)$ does not actually define a connection, unless you regard it as a piece of a Cartan connection. Hence, its exterior derivative – viz., torsion, when the connection is the teleparallelism one – is not really a curvature, except in the Cartan sense again. Indeed, one can speak of $d\,\theta^\mu$ even in the absence of choice of connection, and the significance of $d\,\theta^\mu$ is more related to the integrability of the the vertical sub-bundle of $T(GL(M))$ into the fibers of $GL(M)$ than it is to the torsion of a particular choice of $\mathbf{R}^4$- connection. Hence, one should be wary of whether we are truly defining a translational gauge theory in fact or simply accounting for a missing part of a *Lorentz* gauge theory that relates to appearance of a canonical 1-form on the bundle of Lorentzian frames that does not have an analog in more general gauge theories; i.e., principal fiber bundles that are not frame bundles of some description.

We now address the possible forms that a Lagrangian for $\theta^\mu$ must take (cf., [**11-13**]). In particular, one must look at all of the possible non-trivial 4-forms on $M$ that one can construct from $\mathbf{e}_\mu$, $\theta^\mu$, $g$, $\vartheta$, and $\Theta^\mu$. There are basically three and they are all of the generalized "kinetic energy" form:

$$\eta_{\mu\nu}\Theta^\mu \wedge *^{(\alpha)}\Theta^\nu, \tag{4.11}$$

in which:

$$^{(2)}\Theta^\nu = \tfrac{1}{3}\theta^\nu \wedge \Lambda, \tag{4.12a}$$

$$^{(3)}\Theta^\nu = -\tfrac{1}{3}*(\theta^\nu \wedge *\Phi), \tag{4.12b}$$

$$^{(1)}\Theta^\nu = \Theta^\nu - {}^{(2)}\Theta^\nu - {}^{(3)}\Theta^\nu. \tag{4.12c}$$

In these expressions, we have introduced the *torsion* 1-*form:*

$$\Lambda = i_{\mathbf{e}_\alpha}\Theta^\alpha, \tag{4.13}$$

and the 3-form:

$$\Phi = \eta_{\mu\nu}\Theta^\mu \wedge \theta^\nu, = \eta_{\mu\nu}d\theta^\mu \wedge \theta^\nu, \tag{4.14}$$

which is closely related to what one may call the *Frobenius 3-form* $\theta^\mu \wedge d\theta^\mu$ that is defined by each $\theta^\mu$. This latter 3-form must vanish in order for the exterior differential system that is defined by $\theta^\mu$, namely, $\theta^\mu = 0$, to be integrable into a codimension-one foliation by integral hypersurfaces. Recall that when we regard $\theta^\mu$ as the canonical 1-form on $GL(M)$ the integrability is guaranteed since the integral submanifolds of that exterior differential system are the fibers of $GL(M)$; hence, this invariant would necessarily vanish on $GL(M)$. Indeed, since the coframe field $\theta^\mu$ on $M$ is of maximal rank (i.e., four), its integral submanifolds will be zero-dimensional, namely, the points of $M$. Furthermore, since $\theta^\mu$ is a coframe field one must always some 1-forms $\alpha_\mu$ such that $d\theta^\mu = \alpha_\mu \wedge \theta^\mu$, which makes each $\theta^\mu \wedge d\theta^\mu$ vanish by antisymmetry. The only way that this expression can be possibly non-vanishing is if the number of elements in the coframe $\theta^\mu$ is less than maximal.

Furthermore, one can see that for the present choice of connection the torsion 1-form also takes the form:

$$\Lambda = \text{Tr}\{L_{\mathbf{e}_\alpha}\theta^\beta\}, \tag{4.15}$$

in which L refers to the Lie derivative operator.

*c. Teleparallelism Lagrangian.* One forms the most general teleparallelism Lagrangian from a scalar combination of the aforementioned three parts. In particular, the combination that has generally been used [**12,13**] is:

$$+_{\parallel} = \frac{1}{2l^2}\,\eta_{\mu\nu}\Theta^{\mu} \wedge *\{-^{(1)}\Theta^{\nu} + 2^{(2)}\Theta^{\nu} + \tfrac{1}{3}{}^{(3)}\Theta^{\nu}\}. \tag{4.16}$$

In this expression, $l$ refers to the Planck length. We further introduce the following 2-form with values in $R^4$:

$$H^{\nu} = \frac{1}{2l^2}\{-^{(1)}\Theta^{\nu} + 2^{(2)}\Theta^{\nu} + \tfrac{1}{3}{}^{(3)}\Theta^{\nu}\}, \tag{4.17}$$

and the Euler-Lagrange equations that are associated with varying $+_{\parallel}$ with respect to $\theta^{\mu}$ are ([10]):

$$\delta H_{\mu} = E_{\mu} \tag{4.18}$$

in which we have introduce the energy-momentum 1-form $E_{\mu}$ that is associated with $\theta^{\mu}$ by way of Nöther's theorem:

$$E_{\mu} = \tfrac{1}{2}*(i_{\mathbf{e}_{\mu}}\Theta^{\nu} \wedge *H_{\nu} - \Theta^{\nu} \wedge i_{\mathbf{e}_{\mu}}*H_{\nu}) \tag{4.19}$$

If we wish to include a contribution from a material source distribution then we simply add another term $+_{\text{mat}}$ to the Lagrangian, which adds a further current to the right-hand side by way of $t_{\mu} = \delta +_{\text{mat}}/\delta\theta^{\mu}$. We then rewrite the field equation for $\Theta^{\mu}$ as:

$$\delta H_{\mu} - E_{\mu} = \mathbf{t}_{\mu} \tag{4.20}$$

in order to exhibit the geometrical terms on one side and the material ones on the other.

We note that (4.18) takes the form of the non-trivial Maxwell equation when $H_{\mu}$ corresponds to the electromagnetic induction 2-form that is associated with the field strength 2-form, which is $\Theta^{\mu}$, in this case. Of course, this is to be expected in a gauge field theory, but we are trying to be cautious in or use of that term in the present context.

*d. Equivalence with Einstein's theory.* First, we put the Einstein-Hilbert Lagrangian into the form:

$$+_{\text{EH}} = \frac{1}{2l^2}\,\sqrt{|\det g|}\,\eta_{\mu\nu}\,\tilde{\Omega}^{\mu}_{\lambda} \wedge *(\theta^{\lambda} \wedge \theta^{\nu}), \tag{4.21}$$

and then define the equivalence of the two field theories to mean that under some association of their constituents the field Lagrangians are de Rham cohomologous as 4-forms; i.e., they differ by an exact 4-form (by Stokes's theorem, this would give a vanishing contribution to the field action if $M$ has no boundary).

The identification is effected by going back to the definition of the contortion tensor $\tau = \omega - \varpi$, which we express as

$$\tau_{\mu\nu} = \tfrac{1}{2}(i_{\mathbf{e}_{\nu}}\Theta_{\mu} - i_{\mathbf{e}_{\mu}}\Theta_{\nu} - \eta_{\rho\sigma}\Theta^{\rho}(\mathbf{e}_{\mu}, \mathbf{e}_{\nu}) \wedge \theta^{\sigma}), \tag{4.22}$$

---

[10] The notation on the right-hand side is chosen to be consistent with the literature [**12,13**], despite its conflict with the notation for the basic vector fields on $GL(M)$ that are defined by a linear connection.

in order to show that the Riemannian curvature 2-form $\tilde{\Omega}^{\mu}_{\lambda}$ that is associated with $\varpi = \omega - \tau$ depends essentially only upon the data of the teleparallel geometry since:

$$\tilde{\Omega} = \tilde{\nabla}\,\varpi = \tilde{\nabla}\,\omega - \tilde{\nabla}\,\tau = -\,(d\tau - \tau \wedge \tau + \tau \wedge \omega - \omega \wedge \tau), \qquad (4.23)$$

With these associations, namely, $(\theta^{\mu}, \Theta^{\mu}) \leftrightarrow (\theta^{\mu}, \varpi)$ and $\Theta^{\mu} \leftrightarrow \tilde{\Omega}$, one finds that:

$$+_{EH} - +_{\|} = \frac{1}{l^2} d(\eta_{\mu\nu}\theta^{\mu} \wedge {}^*\Theta^{\mu}), \qquad (4.24)$$

and the Lagrangians are indeed cohomologous.

## 5. Gauge theories on reductions of $GL(M)$.

Now, we recast the previous formalism in terms of the geometry of $GL(M)$. As pointed out above, one advantage of defining ones geometrical objects on $GL(M)$ is that when one chooses a linear connection $\omega$ on $GL(M)$ and a basis for $\mathbf{R}^4$, the basic vector fields $E_{\mu}$ on $GL(M)$ trivialize the horizontal sub-bundle $H(GL(M))$ even when the tangent bundle to $M$ is not trivializable. Hence, the proper context in which to discuss teleparallelism is generally the geometry of $GL(M)$ itself, not the geometry of $M$.

When one reduces $GL(M)$ to a $G$-structure for some subgroup $G$ in $GL(4)$ if $\omega$ is a reducible connection then it takes its values in the Lie algebra $\mathfrak{g}$ that is associated with $G$. The reduction does not change the nature of the horizontal bundle, it merely reduces the vertical sub-bundle, whose tangent spaces become linearly isomorphic to $\mathfrak{g}$ instead of $\mathfrak{gl}(4)$. Hence, the same basic vector fields can still be used; what has changed is set of frames on $M$ that one can deal with and the Lie algebra of infinitesimal transformations that can be applied to the frames.

In the context of gauge theories of spacetime structure our gauge group $G$ is assumed to be a subgroup of $GL(4)$ and our gauge structure is a reduction of $GL(M)$ to a $G$-structure $G(M)$. However, unlike more general gauge theories, we now have to contend with *three* geometrically significant fields: $t$, $\theta^{\mu}$, $\omega$ - not just $\omega$ - and their exterior covariant derivatives $Q$, $\Theta^{\mu}$, $\Omega$, and Bianchi identities. This suggests that, in general, we will need three types of field sources and three differential field equations, possibly of the form (4.20), if not algebraic field equations, such as $Q = 0$.

For instance, the Einstein equations that are associated with an $SO(3,1)$ reduction of $GL(M)$ to a bundle of oriented Lorentz frames with a fundamental tensor field $t = g_{\mu\nu}$, and its Levi-Civita connection $\nabla$, basically make $Q$ and $\Theta^{\mu}$ equal zero and then couple $\Omega$ to the stress-energy-momentum tensor $T$, which serves as source. Of course, this equation can be regarded as a system of differential equations in either $\nabla$ or $g_{\mu\nu}$, or an algebraic equation in $\Omega$, so it does not take the form of a Yang-Mills-type equation.

Teleparallelism was intended to be a gauge theory of gravitation in which $G = \mathbf{R}^4$, $\omega = \theta^{\mu}$, $\Omega = \Theta^{\mu}$. For a "canonical" choice of field Lagrangian, the field equation for $\Theta^{\mu}$, namely, (4.20), took on a Yang-Mills-like form. However, as we have pointed out previously, it is not precisely correct to identify $G$ with $\mathbf{R}^4$. Consequently, we need to reconcile the results of teleparallelism, which seem quite tempting, with the customary practice of gauge field theories.

First, we need to lift the geometrical and physical objects that were previously defined on a parallelizable $M$ to their corresponding analogs on $G(M)$, for some appropriate $G$, when $M$ is not assumed to be parallelizable. The basic prescription is to simply replace any object that is defined on $M$ with a $G$-invariant object that is defined on $G(M)$. For instance, one replaces $T(M)$ with $H(G(M))$ and the vector fields on $M$ with $G$-invariant sections of $H(G(M))$, i.e., $G$-invariant horizontal vector fields on $G(M)$.

Furthermore, the *choice* of coframe field in $M$, namely, $\theta^\mu$, gets replaced with the *canonical* coframe field on $H(G(M))$, which we denote by the same symbol. Hence, there is a subtlety that is associated with interpreting the action of infinitesimal convections on a canonical object. The resolution of this source of confusion is simply the fact that a $G$-reduction of $GL(M)$ is not generally unique, sometimes not even up to homotopy. Hence, a variation of the canonical 1-form of $G(M)$ amounts to a $G$-invariant infinitesimal deformation of the reduction itself within its homotopy class. In effect, this amounts to deforming *all* of the frames in the fibers of $G(M)$ at each point of $M$, not just *one* frame.

One can even speak of deforming the canonical 1-form on $GL(M)$ itself, when there is no space of neighboring reductions in which to deform $GL(M)$. This is because the canonical 1-form is associated with a fundamental tensor field $I: M \rightarrow T^*(M) \otimes T(M)$ on $M$ that takes the form of associating each point $x \in M$ with the identity map on $T_x(M)$, which one regards as an element of $T_x^*(M) \otimes T_x(M)$; for a particular choice of frame $\mathbf{e}_\mu$ in $GL_x(M)$, it takes the form of $\theta^\mu \otimes \mathbf{e}_\nu$. Hence, we can regard a deformation of the canonical 1-form on $GL(M)$ as being equivalent to a deformation of this identity map at each point. It is interesting that we also showed above that the variational formulation of the canonical energy-momentum tensor $T$ on $M$ produced such a deformation, as well. This makes it tempting to set:

$$\delta\theta^\mu = T_\nu^\mu \theta^\nu \qquad (5.1)$$

to begin with. First, we need to re-examine variational field theory for fields that are defined on $G(M)$, in order to make sense of the terms in this equation.

*a. Variational calculus on $G(M)$.* First, we define a physical field to be a $G$-equivariant map $\phi: G(M) \rightarrow N$, where $N$ is a $G$-space, i.e., a manifold on which a left action of $G$ is defined. Considering the fact that $G$ acts on $G(M)$ on the *right*, this means that for every $g \in G$ one must have:

$$\phi(\mathbf{e}g^{-1}) = g\phi(\mathbf{e}). \qquad (5.2)$$

For linear field theories, $N$ is a vector space $V$ and the action of $G$ is linear. Hence, in such an event there is a representation of $G$ in $GL(V)$, which we denote in the customary physical fashion:

$$D: G \rightarrow GL(V), \qquad\qquad g \infty\ D(g), \qquad (5.3)$$

which differentiates at the identity to a representation of the Lie algebra of g in $\mathfrak{gl}(V)$:

$$\mathbf{D}: \mathfrak{g} \to \mathfrak{gl}(V), \qquad a \mapsto \mathbf{D}(a). \qquad (5.4)$$

A particularly useful choice of representative for $V$ is $\Lambda^k(\mathbf{R}^{20}) \otimes \mathbf{R}^M$, where $\mathbf{R}^M$ may also be reducible to a tensor product, since this class of representation includes all scalar, vector, tensor, and spinor fields on $M$. For instance, if:

$$\phi: GL(M) \to \Lambda^k(\mathbf{R}^{20}) \otimes \mathbf{R}^M, \qquad \mathbf{e}_\mu \mapsto \phi^M_{i_1 \mathbf{L} i_k}(\mathbf{e}_\mu) \qquad (5.5)$$

then the associated scalar, vector, etc., field on $M$ is:

$$\phi: M \to \Lambda^k(M) \otimes \mathbf{R}^M, \qquad x \mapsto \frac{1}{k!} \phi^M_{i_1 \mathbf{L} i_k}(\mathbf{e}_\mu)\, \theta^{1 \wedge} \ldots {}^\wedge \theta^k \otimes (\mathbf{e}_\mu \otimes \ldots \otimes \theta^\nu), \qquad (5.6)$$

in which $\mathbf{e}_\mu$ is *any* $G$-frame at $x$ and $\theta^\mu$ is its reciprocal coframe, and the expression $(\mathbf{e}_\mu \otimes \ldots \otimes \theta^\nu)$ depends upon what sort of decomposition one makes of $\mathbf{R}^M$. Although (5.6) looks like a local expression, in fact, the $G$-invariance of $\phi$ makes the expressions well-defined at every point of $M$ without the necessity of defining a global frame field. If we simply abbreviate the vector space $\Lambda^k(\mathbf{R}^{20}) \otimes \mathbf{R}^M$ by $\mathbf{R}^N$ then this $G$-invariance takes the form:

$$\phi(\mathbf{e}g^{-1}) = \mathbf{D}(g)\phi(\mathbf{e}), \qquad \text{for all } \mathbf{e} \in G(M),\, g \in G. \qquad (5.7)$$

Instead of dealing with $J^1(M, \mathbf{R}^N)$, as we did previously, we are now dealing with $J^1(G(M), \mathbf{R}^N)$, which locally looks like $(x^\mu,\ \mathbf{e}_\nu, y^M_{i_1 \mathbf{L} i_k}, p^M_{i_1 \mathbf{L} i_k, \mu}, q^M_{i_1 \mathbf{L} i_k, \nu})$. The 1-jet prolongation of $\phi$ then takes the local form $(x^\mu, \mathbf{e}_\nu, \phi^M_{i_1 \mathbf{L} i_k}(\mathbf{e}_\mu), \partial \phi^M_{i_1 \mathbf{L} i_k}/\partial x^\mu\big|_{\mathbf{e}}, \partial \phi^M_{i_1 \mathbf{L} i_k}/\partial \mathbf{e}_\nu\big|_{\mathbf{e}})$. However, we also need keep in mind that the group $G$ acts on both $G(M)$ and $\mathbf{R}^N$ and that $\phi$ must satisfy (5.7). Hence, we also to specify the manner by which the action of $G$ on $G(M)$ differentiates into an action of $G$ on $J^1(G(M), \mathbf{R}^N)$. From (5.7) and the local action of $G$ on $G(U) = (x^\mu, \mathbf{e}_\nu)$, we see that all we need to resolve is how $G$ acts on $D\phi$.

Choose $g \in G$. Let $R_g: G(M) \to G(M)$ represent right translation by $g$. Since this a diffeomorphism, one has that $DR_g\big|_{\mathbf{e}}$ is a linear isomorphism at any $\mathbf{e} \in G(M)$. Similarly, since the action of $\mathbf{D}(g)$ on $\mathbf{R}^N$ is by a linear isomorphism, its differential is also $\mathbf{D}(g)$. Hence, we can define the action of $g$ on $Df\big|_{\mathbf{e}}$ by considering the commutative diagram:

$$
\begin{CD}
T_{\mathbf{e}} G(M) @>{D\phi|_{\mathbf{e}}}>> \mathbf{R}^N \\
@V{DR_g|_{\mathbf{e}}}VV @VV{D(g)}V \\
T_{\mathbf{e}g^{-1}} G(M) @>{D\phi|_{\mathbf{e}g^{-1}}}>> \mathbf{R}^N
\end{CD}
$$

This gives:

$$gD\phi = D\phi|_{\mathbf{e}g\text{-}1} = D(g)\bullet D\phi\bullet DR_{g\text{-}1}. \tag{5.8}$$

In local form, since $g$ does not depend upon $x$, we can write this as:

$$g\phi^N_{i_1\mathbf{L}i_k,\mu} = \phi^N_{i_1\mathbf{L}i_k,\mu}, \qquad g\phi^N_{i_1\mathbf{L}i_k,\mathbf{e}} = D^N_M(g)\phi^M_{\dot{n}\mathbf{L}i_k,\mathbf{e}}g^{-1}. \tag{5.9}$$

Equation (5.8) allows us to prolong the action of $G$ on $G(M)$ to an action of $G$ on $J^1(G(M), \mathbf{R}^N)$, which we describe in local form:

$$g(x^\mu, \mathbf{e}_\nu, y^N_{\dot{n}\mathbf{L}i_k}, p^N_{i_1\mathbf{L}i_k,\mu}, q^N_{\dot{n}\mathbf{L}i_k,\nu})$$
$$= (x^\mu, \mathbf{e}_\nu g^{-1}, D^N_M(g)y^M_{\dot{n}\mathbf{L}i_k}, p^M_{i_1\mathbf{L}i_k,\mu}, D^N_M(g)q^M_{\dot{n}\mathbf{L}i_k,\nu}g^{-1}). \tag{5.10}$$

It is straightforward to specify that a Lagrangian density on $J^1(G(M), \mathbf{R}^N)$ is a $G$-invariant 4-form that takes the form of $L\mathcal{9}$, where $L$ is a $G$-invariant function on $J^1(G(M), \mathbf{R}^N)$ and $\mathcal{9}$ is the pullback of the volume element on $M$ by the projection of $J^1(G(M), \mathbf{R}^N)$, onto $M$. Furthermore, an allowable variation must take the form of a *G-invariant* vector field on $J^1(G(M), \mathbf{R}^N)$. We next examine the form of such vector fields, keeping in mind that a vector field on $G(M)$ is assumed to be decomposable into a horizontal part and a vertical one:

$$\delta\mathbf{e} = \delta_H\mathbf{e} + \delta_V\mathbf{e}, \tag{5.11}$$

which can be given the local form:

$$\delta\mathbf{e}_\mu = \varepsilon^\mu\partial_\mu - \mathbf{e}_\nu\, a^\nu_\mu. \tag{5.12}$$

We can also decompose $D\phi(\delta\mathbf{e})$ into the sum:

$$D\phi\,(\delta\mathbf{e}) = D\phi(\delta_H\mathbf{e}) + D\phi(\delta_V\mathbf{e}) = \nabla_{\delta\mathbf{e}}\phi + D_V\phi(\delta_V\mathbf{e}), \tag{5.13}$$

so we define:

$$\delta_H\phi = \nabla_{\delta\mathbf{e}}\phi, \qquad \delta_V\phi = D\phi|_{\mathbf{e}}(\delta_V\mathbf{e}) = \mathbf{D}(a)\,\phi(\mathbf{e}). \tag{5.14}$$

The local form of these equations is:

$$\delta_H\phi^{\,N} = \frac{\partial\phi^N}{\partial x^\mu}\varepsilon^\mu + \mathbf{D}^N_M[\Gamma^\lambda_{\mu\nu}\varepsilon^\mu]\phi^M, \qquad \delta_V\phi = \mathbf{D}^N_M(a^\mu_\nu)\,\phi^M, \tag{5.15}$$

in which we have defined:

$$\Gamma^\lambda_{\mu\nu} = \omega^\lambda_\nu(\partial_\mu)\,. \tag{5.16}$$

For any frame $\mathbf{e}_\mu\in G_x(M)$ a horizontal variation $\delta\mathbf{e}_\mu\in H_\mathbf{e}(G_x(M))$ represents the infinitesimal generator of a parallel translation of $\mathbf{e}_\mu$; this amounts to a horizontal lift of an infinitesimal translation (or convection) of the origin of the frame, namely $x\in M$. A vertical variation of $\mathbf{e}_\mu$ then represents an infinitesimal $G$-transformation of $\mathbf{e}_\mu$ that fixes its origin. For instance, when $G = SO(3,1)$, this would be an infinitesimal Lorentz transformation.

Now, we need to examine how the variations $\delta\mathbf{e}$ and $\delta\phi$ differentiate to a $G$-invariant variation of $D\phi$. Going back to the commutative diagram, we see that this simply amounts to differentiating the action of $G$ at the identity, which is $\mathfrak{g}$. If $a \in G$ and $\mathbf{e} \in G(M)$ then the infinitesimal form of the diagram is now:

$$
\begin{array}{ccc}
T_\mathbf{e}G(M) & \xrightarrow{\ D\phi|_\mathbf{e}\ } & \mathbf{R}^N \\[2mm]
\Big\downarrow{\scriptstyle R_a|_\mathbf{e}} & & \Big\downarrow{\scriptstyle \mathbf{D}(a)} \\[2mm]
T_\mathbf{e}G(M) & \xrightarrow{\ D\phi|_\mathbf{e}\ } & \mathbf{R}^N
\end{array}
$$

From the linearity of $D\phi|_\mathbf{e}$ and the linearity of the action of $\mathfrak{g}$ on $\mathbf{R}^N$ by way of $\mathbf{D}$, we see that we can set:

$$\delta D\phi = \delta_H(D\phi) + \delta_V(D\phi)$$
$$= D(\delta_H\phi) - \mathbf{D}(a)D\phi(\mathbf{e}a) = \nabla(\delta\phi) - \mathbf{D}(a)D\phi(\mathbf{e}a). \tag{5.17}$$

Suppose $\phi$ is a $k$-form on $G(M)$ with values in $\mathbf{R}^N$. Let:

$$\mathbf{v} = \delta x^\mu + \delta\mathbf{e}_\mu \tag{5.18}$$

be a $G$-invariant vector field on $G(M)$. Its 1-jet prolongation is:

$$\hat{\mathbf{v}} = \delta x^\mu + \delta\mathbf{e}_\mu + \delta_H\phi + \delta_V\phi + \delta_H\phi_{,\mu} + \delta_V\phi_{,\mu}, \tag{5.19}$$

which we rewrite in "horizontal + vertical form" as:

$$\hat{\mathbf{v}}_H + \hat{\mathbf{v}}_V = (\varepsilon^\mu\partial_\mu + \varepsilon^\mu\nabla_\mu\phi\,) + (-\mathbf{e}_\mu a + \mathbf{D}(a)\phi - \mathbf{D}(a)D\phi(\mathbf{e}a)), \tag{5.20}$$

i.e.:

$$\delta x^\mu \ = \varepsilon^\mu\partial_\mu, \qquad\qquad \delta\mathbf{e}_\mu \ = -\mathbf{e}_\varpi\, a^\nu_\mu, \tag{5.21a}$$

$$\delta_H\phi^N = 0, \qquad\qquad \delta_V\phi^N = \mathbf{D}^N_M(a)\phi^M\,, \tag{5.21b}$$

$$\delta_H\phi^N_{,\mu} = \varepsilon^\mu\nabla_\mu\phi^N, \qquad\qquad \delta_V\phi^N_{,\mu} = \mathbf{D}^N_M(a)\phi^M_{,\mu}(\mathbf{e}_\mu a)\,, \tag{5.21c}$$

For specificity, suppose $\phi$ is a $G$-invariant $p$-form on $G(M)$ that takes its values in $\mathbf{R}^N$. The total variation of $+$ under the prolonged infinitesimal action of $\mathbf{v}$ then takes the local form:

$$\delta + = i_{\hat{\mathbf{v}}}\left(\frac{\partial +}{\partial x^\mu}dx^\mu + \frac{\partial +}{\partial\mathbf{e}_\mu}d\mathbf{e}_\mu + \frac{\partial +}{\partial\phi}{}^\wedge d\phi + \frac{\partial +}{\partial(D\phi)}{}^\wedge d(D\phi)\right)\boldsymbol{\vartheta} + di_{\hat{\mathbf{v}}}+$$

$$= \left[\frac{\partial +}{\partial x^\mu}\varepsilon^\mu + \frac{\delta +}{\delta\phi}{}^\wedge\varepsilon^\mu\nabla_\mu\phi\,\right] - \left[\frac{\partial +}{\partial\mathbf{e}_\mu}(\mathbf{e}_\mu a) - \frac{\delta +}{\delta\phi}{}^\wedge\mathbf{D}(a)\,\phi(\mathbf{e})\right]$$

$$+ d\!\left((-1)^{4-p-1}\frac{\partial +}{\partial\phi_{,\mu}}{}^\wedge(\varepsilon^\nu\nabla_\mu\phi + \mathbf{D}(a)\phi) - Li_{\delta x}\,\boldsymbol{\vartheta}\right). \tag{5.22}$$

Hence, the least action principal says that when $\phi$ is an extremal field the variation of the action defined by $+$ should vanish for any $G$-invariant vertical variation of the frames in $G(M)$. One still writes the resulting Euler-Lagrange equation for $\phi$ as:

$$\frac{\delta +}{\delta \phi} = 0. \tag{5.23}$$

The Nöther current associated with the $G$-invariant variation $\mathbf{v} = \varepsilon^\mu E_\mu + a^k E_k$ is a $G$-invariant vector field on $G(M)$:

$$\mathbf{J} = (T_\nu^\mu \varepsilon^\nu) E_\mu + (\mathbf{D}^k(a)\phi) E_k, \tag{5.24}$$

with:

$$T_\nu^\mu = \frac{\partial +}{\partial \nabla_\mu \phi} \nabla_\nu \phi - L \delta_\nu^\mu. \tag{5.25}$$

In particular, when $\mathbf{v}$ is a horizontal variation, the current is energy-momentum, and when $\delta \mathbf{e}_\mu$ is a vertical variation the current is $G$-momentum, such as angular momentum when $G$ is an orthogonal subgroup of $GL(4)$.

*b. Teleparallelism on $G(M)$.* We can now reformulate teleparallelism in terms of geometrical objects on $G(M)$. The immediate difference is that instead of making our specific choice of $\phi$ in the form of a global frame on $M$, which might not exist, we deal with the canonical 1-form $\theta^\mu$ on $G(M)$, which always does. Hence, this makes $\mathbf{R}^N = \mathbf{R}^4$, and $\mathrm{D}(g)$ becomes the representation of $g$ by an invertible $4 \times 4$ real matrix: if $\mathbf{e}_\mu$ goes to $\mathbf{f}_\mu = \mathbf{e}_\mu g^{-1}$ then $g$ goes to $\mathrm{D}(g) = g_\nu^\mu$, which represents the matrix of components of the frame $\mathbf{f}_\mu$ with respect to $\mathbf{e}_\mu$; its action on $\theta^\mu$ is then simply:

$$\mathrm{D}(g)\,\theta^\mu = g_\nu^\mu \, \theta^\nu. \tag{5.26}$$

Similarly, the variations that we will use are not the infinitesimal convections $\delta x^\mu$ on $M$ but the $G$-invariant horizontal vector fields $\delta_H \mathbf{e}$ on $G(M)$ – i.e., infinitesimal parallel translations – that represent the horizontal lifts of vector fields on $M$. The variation of $\theta^\mu$ that such a vector field generates is then:

$$\delta\theta^\mu = \mathrm{L}_{\delta_H \mathbf{e}}\theta^\mu = d(\delta_H \mathbf{e})^\mu + i_{\delta_H \mathbf{e}}d\theta^\mu = i_{\delta_H \mathbf{e}}\Theta^\mu - i_{\delta_H \mathbf{e}}(\omega \wedge \omega)$$
$$= i_{\delta_H \mathbf{e}}\Theta^\mu. \tag{5.27}$$

An interesting subtlety to note is that if we had simply *defined* $\Theta^\mu$ to be $d\theta^\mu$ then we would have arrived at the same result for $\delta\theta^\mu$, independently of the choice of $\omega$, except insofar as we used it to define horizontal variations. In some formulations of teleparallelism (cf. [**12**]) the role of the connection that defines $\Theta^\mu$ is trivialized by setting $\omega = 0$ anyhow. This begs the deeper question of whether teleparallelism has more to do with accounting for the contribution that $\theta^\mu$ and $d\theta^\mu$ make to a physical field theory than it does with accounting for the geometry of global frame fields under the induced connections.

A Lagrangian density for $\theta^\mu$ then becomes a $G$-invariant 4-form on $J^1(G(M), \mathbf{R}^4)$ of the form:

$$+ = L(x^\mu, \mathbf{e}_\mu, \theta^\mu, d\theta^\mu) \, \boldsymbol{\vartheta}. \tag{5.28}$$

In particular, if we want to borrow the previous Lagrangian that we used in the case of a parallelizable $M$:

$$+_\| = \frac{1}{2l^2} \, \eta_{\mu\nu} \Theta^\mu \wedge *\{-{}^{(1)}\Theta^\nu + 2{}^{(2)}\Theta^\nu + \tfrac{1}{3}{}^{(3)}\Theta^\nu\}, \tag{5.29}$$

then we must notice that the $G$ for which this expression is well-defined is the Lorentz group; i.e., $G(M)$ must be the bundle of oriented Lorentzian frames. Otherwise, if we wished to define this Lagrangian for $G = SL(4)$ then we would have to use a metric that was non-constant on the more general frames, and this would imply extra terms in the field equations and conserved currents to account for the differentiation of the metric tensor.

The main difference between the previous formulation of teleparallelism on $M$ and the present one is that now the various $\mathbf{R}^4$-valued 2-forms are defined on $SO(3,1)(M)$, not on $M$. However, in order to emphasize the fact that we have essentially "lifted" $T(M)$ to $H(SO(3,1)(M))$, we point out that the geometrical objects on $M$ that we defined before must be associated with the restrictions of the corresponding objects on $SO(3,1)(M)$ to the horizontal sub-bundle. In most cases, this is merely a case of replacing derivatives with covariant derivatives, which involve first projecting into the horizontal sub-bundle. However, one must still be cautious about whether some operations on horizontal objects may imply a vertical part as a by-product. We summarize the key associations as follows:

$$
\begin{aligned}
M &\rightarrow SO(3,1)(M) \\
T(M) &\rightarrow H(SO(3,1)(M)) \\
\mathrm{X}(M) &\rightarrow SO(3,1)\text{-invariant horizontal vector fields on } SO(3,1)(M) \\
\Lambda^k(M) &\rightarrow SO(3,1)\text{-invariant horizontal } k\text{-forms on } SO(3,1)(M) \\
\mathbf{e}_\mu &\rightarrow E_\mu \\
\theta^\mu &\rightarrow \theta^\mu \\
D\mathbf{e}_\mu &\rightarrow \nabla E_\mu \\
D\theta^\mu &\rightarrow \nabla \theta^\mu \\
d\theta^\mu &\rightarrow \Theta^\mu.
\end{aligned}
$$

The field equations for $\theta^\mu$ are formally unchanged from the previous ones, viz.:

$$\delta H_\mu - E_\mu = \mathbf{t}_\mu \tag{5.30}$$

with:

$$H^\nu = \frac{1}{2l^2}\{-{}^{(1)}\Theta^\nu + 2{}^{(2)}\Theta^\nu + \tfrac{1}{3}{}^{(3)}\Theta^\nu\}, \tag{5.31a}$$

$$E_\mu = \tfrac{1}{2}*(i_{E_\mu}\Theta^\nu \wedge *H_\nu - \Theta^\nu \wedge i_{E_\mu}*H_\nu). \tag{5.31b}$$

However, this time, in order to make sense of the *-operator, we simply use a Lorentz-invariant volume element on the bundle $H(SO(3,1)(M))$ to associate horizontal $k$-vector fields with horizontal 4-$k$-forms. However, one notes that the definition $\delta = *^{-1}d*$ includes the possibility that the exterior derivative of the horizontal 4-$k$-form might still have a vertical part. Hence, we replace it with the definition:

$$\delta = *^{-1}\nabla*. \tag{5.32}$$

**6. Discussion.** The concluding remark of the last section underscores something about teleparallelism that becomes self-evident when one formulates it on $SO(3,1)(M)$: it is not a complete statement of the geometry of spacetime, much less how that geometry is coupled to the field theories of physics. It is essentially a statement of how the geometry of the horizontal sub-bundle of $T(GL(M))$ for a choice of linear connection is coupled to the energy-momentum current of a matter distribution in a manner that accounts for the presence of gravitation as manifestation of the *torsion* of the connection. In particular, one must still account for the vertical part of geometry – viz., $\omega$ and $\Omega$ – and how it relates to physics.

However, if one simply accepts this purely horizontal aspect of the theory then it still has something intriguing to say about physics and spacetime structure: Just as $\theta^\mu$ and $d\theta^\mu$ represent canonical structures on $GL(M)$, similarly, gravitation has a sort of canonical character in physics. By this, we mean that it is universal in its effects and seems to represent a residual interaction that exists when the effects of the all of the other three fundamental interactions have averaged out. Consequently, even though gravitation is, by many orders of magnitude, the weakest of the four interactions, nevertheless, it is the one that ultimately dominates on the astronomical scale. This would seem to be consistent with the notion that a reduction of $GL(4)$ to $\{e\}$ is the furthest that one can go, and the associated reduction of $GL(M)$ is to a global frame field.

However, as pointed out above, this possibility should be approached with caution. A more prudent approach would be to look at maximal partial sections:

$$\mathbf{e}: \&[\mathbf{e}] \to GL(M), \tag{6.1}$$

whose domains $\&[\mathbf{e}]$ are complementary to the carriers $\mho[\mathbf{e}]$ of obstruction cycles. One would thus define the generic and singular points of such partial frame fields. One could then define a partial connection $({}^{11})$ on $GL(M)$ by pushing the maximal section $\mathbf{e}$ up to a partial frame field on $GL(M)$ by way of:

---

[11] This usage of the term "partial connection" is not consistent with the one that is given in Kamber and Tondeur [**16**], in which the partiality refers to the rank of the horizontal sub-bundle, not its domain of definition.

$$D\mathbf{e}: T(\&[\mathbf{e}]) \rightarrow T(GL(M)), \tag{6.2}$$

and then extending this partial frame field to the other frames of $GL(\&[\mathbf{e}])$ by right-translation.

This still leaves the fibers of singular points unaccounted for. Here, we note a subtlety: Although it is, by definition, impossible to extend $\mathbf{e}$ to its singular set, nevertheless, since global frame fields exist on $GL(M)$, it might be possible to extend the partial frame field on $GL(M)$. The difference is that whereas the partial frame field also spans a partial sub-bundle of $T(GL(M))$ that one could define to be the horizontal complement to the vertical sub-bundle, at least for the generic fibers, nevertheless, one expects that the extended frame field might necessarily 'twist' into the vertical subspaces to some degree, at least for the singular fibers. One naturally wonders how the topological information that is contained in the singularity complex of $\mathbf{e}$ and the topology of $M$ might be used to effect this extension mathematically. The application to physics would undoubtedly involve the association of some sort of constitutive law to account for the non-qualitative aspects of the association of the topology of singular frame fields with the geometry of manifolds.

If one actually wishes to attribute the existence of gravitation to the geometry of an $\{e\}$-reduction of $GL(M)$ – if only a maximal partial one – then one must immediately address the question of the role that is to be played by the spacetime metric tensor field. If one recalls that the true historical origin of relativity theory was in the theory of electromagnetism, in which the metric tensor is closely related to the propagation of electromagnetic waves, then one suspects that perhaps the role of the metric is more closely associated with the electromagnetic structure, after all. Indeed, one can obtain a conformal class of Lorentzian metrics as a consequence of assuming a linear constitutive law for the electromagnetic field strength 2-form (cf. [**14,15**]). Oc course, one would also like to attribute this 2-form to some intrinsic geometrical or topological structure that is associated with the spacetime manifold in its own right, and not merely introduce it *a priori*.